\documentclass[12pt]{article}
\usepackage{amsfonts}
\usepackage{latexsym}
\usepackage{amsmath}
\usepackage{amssymb}
\usepackage{mathrsfs}

\allowdisplaybreaks

\newtheorem{theorem}{Theorem}[section]
\newtheorem{proposition}[theorem]{Proposition}
\newtheorem{corollary}[theorem]{Corollary}

\newtheorem{remark}[theorem]{Remark}

\newcommand{\newsection}{\setcounter{equation}{0}\section}
\def\appendix#1{\addtocounter{section}{1}\setcounter{equation}{0}
\renewcommand{\thesection}{\Alph{section}}
\section*{Appendix \thesection\protect\indent \parbox[t]{11.15cm}{#1}}
\addcontentsline{toc}{section}{Appendix \thesection\ \ \ #1}}

\newcommand{\be}{\begin{eqnarray}}
\newcommand{\ee}{\end{eqnarray}}
\newcommand{\bea}{\begin{eqnarray}}
\newcommand{\eea}{\end{eqnarray}}
\newcommand{\ba}{\begin{array}}
\newcommand{\ea}{\end{array}}

\newcommand{\Ker}{\operatorname{Ker}}

\newenvironment{frcseries}{\fontfamily{frc}\selectfont}{}

\def \la {\label}

\def\e{\epsilon}

\def\bbe{{\bf{e}}}
\font\mybb=msbm10 at 11pt

\def\bb#1{\hbox{\mybb#1}}

\def\bR {\bb{R}}

\def\tn {{\tilde{\nabla}}}

\begin{document}

\title{Supersymmetric near-horizon geometries in $D=6$ supergravity: Lichnerowicz theorems, index theory and symmetry enhancement}
\author{U.~Kayani}
\date{}
\maketitle

\begin{abstract}
We analyse supersymmetric near-horizon geometries of extremal black holes in
$N=(1,0)$, $D=6$ supergravity with one tensor multiplet and $U(1)$ $R$-symmetry gauging.
Assuming smooth bosonic fields and a compact, connected, boundaryless spatial horizon section
$\mathcal{S}$, we solve the Killing spinor equations (KSEs) along the lightcone directions and
identify the independent horizon system satisfied by the spinors $\eta_\pm$ on $\mathcal{S}$.
We then prove generalized Lichnerowicz-type theorems for both lightcone chiralities, showing
that the zero modes of the relevant horizon Dirac operators are in one-to-one correspondence
with Killing spinors on $\mathcal{S}$.

As a consequence, the supersymmetry-counting formula
$N = 2N_{-} + \mathrm{Index}(D_E)$
holds for the class of regular horizons under consideration, where $D_E$ is the horizon Dirac
operator twisted by the bundle naturally associated to the gauge structure of the theory. The
$D=6$ case is distinguished from the previously analysed $D=11$ and type-IIA horizons because
$\mathcal{S}$ is a compact four-manifold and the theory is chiral, so the relevant index need
not vanish. In the ungauged case this reduces to the ordinary chiral Dirac index on
$\mathcal{S}$, while in the gauged case the index is that of the corresponding twisted operator.

We also analyse the map $\eta_- \mapsto \Gamma_+\Theta_-\eta_-$. For non-trivial fluxes, the
resulting spacetime $\mathfrak{sl}(2,\mathbb{R})$ symmetry is proved unconditionally in the
ungauged theory. In the gauged theory the same conclusion follows provided one assumes
$\mathrm{Ker}\,\Theta_- = \{0\}$. We state this assumption explicitly and do not claim a full
gauged symmetry-enhancement theorem without it.
\end{abstract}


\newsection{Introduction}

Supersymmetric near-horizon geometries provide a natural arena in which to analyse extremal black
holes in supergravity. The near-horizon limit isolates the local geometry of a degenerate Killing
horizon while preserving the full system of field equations and Killing spinor equations (KSEs). In
many supergravity theories one finds that supersymmetry near the horizon is larger than in the
corresponding bulk black-hole solution and that the bosonic isometry algebra is enhanced by an
$\mathfrak{sl}(2,\mathbb R)$ factor. This picture underlies the horizon conjecture and is closely
related to attractor behaviour and to the classification of extremal near-horizon geometries
\cite{attract1,astef,nhsymmetrythm1,u1proof,Kunduri:2006ek,Kunduri:2007qy,KL2013review}.

For smooth supersymmetric near-horizon geometries with compact, connected, boundaryless spatial
section $\mathcal S$, the horizon conjecture predicts that
\begin{eqnarray}
N = 2N_{-} + \mathrm{Index}(D_{E})~,
\label{indexcon}
\label{index}
\end{eqnarray}
where $N$ is the total number of Killing spinors, $N_->0$, and $D_E$ is the horizon Dirac operator
twisted by the bundle appropriate to the gauge sector of the theory. It further predicts that if
the fluxes are non-trivial and $N_-\neq 0$, then the near-horizon spacetime admits an
$\mathfrak{sl}(2,\mathbb R)$ isometry subalgebra. This programme has been completed in a number of
supergravity theories, including $D=11$ supergravity, type IIA, massive IIA, type IIB, $D=5$
gauged supergravity with vector multiplets, and $D=4$ gauged supergravity
\cite{11index,iiaindex,miiaindex,iibindex,5dindex,4dindex}.

In this paper we revisit the conjecture for $N=(1,0)$, $D=6$ supergravity with one tensor
multiplet and $U(1)$ $R$-symmetry gauging, i.e. the Salam--Sezgin model and its ungauged limit
\cite{6d1,6d2,6d6,6d14}. The six-dimensional case is qualitatively different from the previously
analysed $D=11$ and type-IIA theories. In those examples the relevant horizon index vanishes: for
$D=11$ because the horizon section is odd-dimensional, and for type IIA because the horizon Dirac
operator acts on non-chiral Majorana spinors \cite{11index,iiaindex,miiaindex}. In the present
$D=6$ theory the horizon section is a compact four-manifold and the supersymmetry parameter is
chiral, so the horizon Dirac operator $\mathscr D^{(+)}$ has a potentially non-zero index. This is
the main structural novelty of the six-dimensional analysis and is the reason that the final
supersymmetry-counting formula is not of the simple form $N=2N_-$. 

There is already substantial six-dimensional literature with which the present analysis must be
compared. Supersymmetric solutions of minimal ungauged six-dimensional supergravity were classified
in \cite{6d15}; near-horizon geometries of $(1,0)$ theories with tensor and hypermultiplets were
analysed in \cite{6d16}; the tensor-multiplet sector without hypermultiplets was revisited in
\cite{6d23}; general supersymmetric solutions of $U(1)$ and $SU(2)$ gauged six-dimensional
supergravities were described in \cite{6d18}; while related horizon and spinorial analyses in six
dimensions can be found in \cite{6d16,6d18,6d23}. Our aim is different from a local classification.
We instead perform a global analysis of the horizon KSEs tailored to
the horizon conjecture, with particular emphasis on separating unconditional statements from those
which remain conditional in the gauged theory.

The main results of the paper are the following. First, after solving the KSEs along the lightcone
directions, we identify the independent differential and algebraic conditions on the horizon spinors
$\eta_\pm$ on $\mathcal S$. Second, we prove generalized Lichnerowicz-type theorems for both
lightcone chiralities, showing that the kernels of the horizon Dirac operators
$\mathscr D^{(\pm)}$ are in one-to-one correspondence with Killing spinors on $\mathcal S$.
Third, we obtain the unconditional supersymmetry-counting formula
\begin{eqnarray}
N = 2N_- + \mathrm{Index}(\mathscr D^{(+)})~,
\end{eqnarray}
where $\mathscr D^{(+)}$ is the positive-chirality horizon Dirac operator defined explicitly in
section~5. In the ungauged theory ($g=0$) the Atiyah--Singer theorem gives
\begin{eqnarray}
\mathrm{Index}(\mathscr D^{(+)}) = -\frac{\mathrm{sign}(\mathcal S)}{8}~,
\end{eqnarray}
so that $N = 2N_- - \mathrm{sign}(\mathcal S)/8$. Since $\mathcal{S}$ is spin (as required for
the horizon spinors to exist), Rokhlin's theorem gives $\mathrm{sign}(\mathcal{S}) = 16k$ for
some $k\in\mathbb{Z}$, so the index equals $-2k$ and
\begin{eqnarray}
N = 2(N_- - k)~, \qquad k = \tfrac{1}{16}\,\mathrm{sign}(\mathcal{S}) \in \mathbb{Z}~,
\end{eqnarray}
is manifestly even. In the gauged theory the index depends on the precise $U(1)$ twisting of $\mathscr D^{(+)}$;
its evaluation is discussed in section~\ref{sec:index}, where we also show that the index is an
integer by the even intersection form on the spin manifold $\mathcal{S}$, and give a sufficient
condition for it to be even. This is the first
example in this series of horizon-conjecture analyses in which the index contribution is
generically non-vanishing.

The status of the symmetry-enhancement statement requires more care. We analyse the map
$\eta_-\mapsto \Gamma_+\Theta_-\eta_-$ and show that in the ungauged theory, if the fluxes are
non-trivial and $N_-\neq 0$, then a maximum-principle argument implies
$\mathrm{Ker}\,\Theta_- = \{0\}$, so the spacetime admits an $\mathfrak{sl}(2,\mathbb R)$
isometry algebra. In the gauged theory the same argument is obstructed by the negative gauging term
in equation~(\ref{nrm1ab}). Accordingly, we do not claim an unconditional proof of the second part
of the horizon conjecture in the gauged case. Instead, we show that the
$\mathfrak{sl}(2,\mathbb R)$ conclusion follows provided one assumes
$\mathrm{Ker}\,\Theta_- = \{0\}$.

We also improve on a common assumption in the near-horizon literature. Several earlier analyses
identify the stationary Killing vector of the black hole with a Killing-spinor bilinear from the
outset; see for example \cite{gutbh,reall,bilin,6d15}. We do not impose this bilinear matching
condition. Rather, it emerges from the solution of the KSEs. Throughout, we assume that the event
horizon is a Killing horizon, so that Gaussian null coordinates can be introduced in a neighbourhood
of the horizon \cite{isen,gnull,rigidity1,axi1,axi2}. Compactness of $\mathcal S$ is used
essentially in the maximum-principle arguments, in the integration-by-parts identities entering the
Lichnerowicz theorems, and in the application of the Atiyah--Singer index theorem.

The paper is organized as follows. In section~2 we review the relevant $N=(1,0)$, $D=6$ theory and
its field equations and KSEs. In section~3 we introduce the near-horizon fields and solve the KSEs
along the lightcone directions. Section~4 identifies the independent KSEs on $\mathcal S$ and
records how the remaining conditions follow from the horizon Bianchi identities and field equations.
In section~5 we prove the Lichnerowicz theorems. Section~6 contains the supersymmetry-counting
result, the explicit index computation, and the symmetry-enhancement analysis, with the gauged and
ungauged cases carefully separated. The appendices summarize the spinor conventions, the near-horizon
spin connection and curvature, and the independent horizon field equations and Bianchi identities.

\newsection{$N=(1,0)$, $D=6$ gauged supergravity}

We review the $N=(1,0)$, $D=6$ gauged supergravity of \cite{6d6, 6d14}. This is a chiral theory with 8 real supersymmetries and $U(1)$ $R$-symmetry gauging. The fermions carry the doublet index of the $R$-symmetry group $Sp(1)_R$ and are all chiral: $\Gamma_{*} \lambda = \pm \lambda$ where $\Gamma_{*} = \Gamma_0 \cdots \Gamma_5$. We take the plus sign throughout and consider left-handed spinors. We have the following multiplets,
\bea
(e_{M}{}^a, \psi_{M}, B^{+}_{M N})&& \, \, \, \, \text{graviton}
\nonumber \\
(\Phi, \chi, B^{-}_{M N})&& \, \, \, \, \text{tensor/dilaton}
\nonumber \\
(A_M, \lambda)&& \, \, \, \, \text{$U(1)$-vector}
\eea
where $B^{\pm}$ gives rise to self-dual/anti-self-dual 3-form field strengths. $\lambda, \chi$ are spin-$\frac{1}{2}$ particles, $\psi_M$ is the spin-$\frac{3}{2}$ gravitino, $A_M$ is the vector gauge field from the $U(1)$ symmetry and $\Phi$ is a dilaton. The Lagrangian is given by,
\bea
{\cal{L}} &=& R \star 1 - \tfrac{1}{4}\star d\Phi \wedge d\Phi -\tfrac{1}{2}e^{\Phi}H_{(3)}\wedge H_{(3)}
\nonumber\\
&-& \tfrac{1}{2}e^{\frac{\Phi}{2}} \star F_{(2)} \wedge F_{(2)} - 8g^2 e^{-\frac{\Phi}{2}} \star 1
\eea
The field strengths $F_{(2)}$ and $H_{(3)}$ are defined by,
\bea
F_{(2)} &=& dA_{(1)}
\nonumber \\
H_{(3)} &=& dB_{(2)} + \frac{1}{2}F_{(2)} \wedge A_{(1)}
\eea
These give rise to the Bianchi identities $dF_{(2)} = 0$ and $dH_{(3)} = \frac{1}{2}F_{(2)}\wedge F_{(2)}$ which in coordinates can be expressed as,
\bea
BF_{M N P} &\equiv& \nabla_{[M}{F_{N P]}} = 0
\nonumber \\
BH_{M N P Q} &\equiv&  \nabla_{[M}{H_{N P Q]}} - \frac{3}{4}F_{[M N}F_{P Q]} = 0
\eea
The field equations for the bosonic fields are as follows. The Einstein equation is
\bea
\label{eins}
E_{M N} &\equiv& R_{M N} - \frac{1}{4}\nabla_{M}{\Phi}\nabla_{N}{\Phi}
- \frac{1}{2}e^{\frac{\Phi}{2}}\bigg(F_{M P}F_{N}{}^{P} - \frac{1}{8}F^2 g_{M N}\bigg)
\nonumber \\
&-& \frac{1}{4}e^{\Phi}\bigg(H_{M P Q}H_{N}{}^{P Q} - \frac{1}{6}H^2 g_{M N}\bigg)
- 2g^2 e^{-\frac{\Phi}{2}} g_{M N} = 0
\eea
The dilaton field equation,
\bea
\label{scalarfeq}
F\Phi \equiv \nabla^{M}{\nabla_{M}}{\Phi} - \frac{1}{4}e^{\frac{\Phi}{2}}F^2 - \frac{1}{6}e^{\Phi}H^2 + 8g^2 e^{-\frac{\Phi}{2}} = 0
\eea
and the field equations for the fluxes,
\bea
\label{2feq}
d(e^{\frac{\Phi}{2}} \star F_{(2)}) &=& e^{\Phi}\star H_{(3)} \wedge F_{(2)}
\eea
\bea
\label{3feq}
d(e^{\Phi}\star H_{(3)}) &=& 0
\eea
In coordinates these can be expressed as,
\bea
FH_{M N} &\equiv& \nabla^{P}{H_{M N P}} + H_{M N P}\nabla^{P}{\Phi} = 0
\nonumber \\
FF_{M} &\equiv& \nabla^{N}{F_{M N}} + \frac{1}{2}F_{M N}\nabla^{N}{\Phi} + \frac{1}{2}F^{N P}H_{M N P} = 0
\eea
The KSEs are given as the vanishing of the supersymmetry transformations of the fermionic fields,
\bea
\label{Gkse}
\delta \psi_{M} \equiv {\cal D}_{M}\epsilon &=& \bigg(\nabla_{M} - i g A_{M} + \frac{1}{48}e^{\frac{\Phi}{2}}H^+_{N P Q}\Gamma^{N P Q}\Gamma_{M}\bigg)\epsilon = 0
\eea
\bea
\label{Akse}
\delta \chi \equiv {\cal A} \epsilon &=& \bigg(\Gamma^{N}\nabla_{N}{\Phi} - \frac{1}{6}e^{\frac{\Phi}{2}} H^-_{N P Q}\Gamma^{N P Q}\bigg)\epsilon = 0
\eea
\bea
\label{Fkse}
\delta \lambda \equiv {\cal F} \epsilon &=& \bigg(e^{\frac{\Phi}{4}} F_{N M}\Gamma^{N M} - 8i g e^{-\frac{\Phi}{4}}\bigg)\epsilon = 0
\eea
where $\epsilon$ is the supersymmetry parameter which from now on is taken to be a commuting symplectic Majorana-Weyl spinor of $Spin(5, 1)$\footnote{$\epsilon$ also has an $Sp(1)$ index which we will suppress}. Note that the $\pm$ superscripts appearing on the 3-form $H_{NPQ}$ in these expressions are redundant, since the
chirality of $\epsilon$ already implies projections onto the self-dual or anti-self-dual parts. The integrability conditions of the KSEs are given by,
\bea
\Gamma^{N}[{\cal D}_{M}, {\cal D}_{N}]\epsilon + \mu_{M} {\cal A}\epsilon + \lambda_{M} {\cal F}\epsilon &=& \bigg( \frac{1}{2}E_{M N}\Gamma^{N} + \frac{1}{12}e^{\frac{\Phi}{2}}BH_{M N P Q}\Gamma^{N P Q}
\nonumber \\
&-&\frac{1}{48}e^{\frac{\Phi}{2}}BH_{N P Q R}\Gamma_{M}{}^{N P Q R} + \frac{1}{8}e^{\frac{\Phi}{2}}FH_{M N}\Gamma^{N}
\nonumber \\
&-& \frac{1}{16}e^{\frac{\Phi}{2}}FH_{N P}\Gamma_{M}{}^{N P} \bigg)\epsilon
\eea
where,
\bea
\mu_{M} &=& \frac{1}{8}\nabla_{M}{\Phi} + \frac{1}{96}e^{\frac{\Phi}{2}}H_{N P Q}\Gamma^{N P Q}\Gamma_{M}
\nonumber \\
\lambda_{M} &=& \frac{1}{64}e^{\frac{\Phi}{4}}F_{N P}\Gamma_{M}\Gamma^{N P} - \frac{1}{8}e^{\frac{\Phi}{4}}F_{M N}\Gamma^{N} + \frac{i}{8}e^{-\frac{\Phi}{4}}g \Gamma_{M}
\eea
we see that if the $H$ field equation, Bianchi identity and the Killing
spinor conditions are satisfied, and given that the Ricci tensor is diagonal, the Einstein equation is then satisfied as well. Additional integrability conditions may be derived from the algebraic conditions as follows,
\bea
\Gamma^{M}[{\cal D}_{M}, {\cal A}]\epsilon + \lambda {\cal A}\epsilon + \mu {\cal F}\epsilon &=& \bigg(F\Phi - \tfrac{1}{6}e^{\frac{\Phi}{2}}BH_{M N P Q}\Gamma^{M N P Q}
\nonumber\\&&
- \tfrac{1}{2}e^{\frac{\Phi}{2}}FH_{N P}\Gamma^{N P}\bigg)\epsilon
\nonumber \\
\Gamma^{M}[{\cal D}_{M}, {\cal F}]\epsilon - \lambda {\cal F}\epsilon -2\mu {\cal A}\epsilon &=& \bigg(e^{\frac{\Phi}{4}}BF_{M N P}\Gamma^{M N P} - 2e^{\frac{\Phi}{4}}FF_{M} \Gamma^{M}\bigg)\epsilon
\eea
where
\bea
\lambda &=& -\frac{1}{24}e^{\frac{\Phi}{2}}H_{M N P}\Gamma^{M N P}
\nonumber \\
\mu &=& \frac{1}{8}e^{\frac{\Phi}{4}} F_{M N}\Gamma^{M N} + i e^{-\frac{\Phi}{4}}g
\eea
The first shows once the $H$ field equation and Bianchi identity and the Killing spinor conditions are satisfied, then the dilaton field equation is satisfied as well. The second is automatically satisfied as a result of the $F$ field equation and the Killing spinor equations.

\newsection{Near-horizon Data and Solution to the KSEs}

To analyse near-horizon geometries we introduce coordinates regular and adapted to the horizon. We consider a six-dimensional stationary black hole metric for which the horizon is a Killing horizon and the metric is regular there. A set of Gaussian Null coordinates \cite{isen, gnull} $\{u, r, y^{i}\}$
will be used to describe the metric, where $r$ denotes the radial distance away from the event horizon which is located at $r=0$ and $y^i,~ i=1, \dots, 4$ are local co-ordinates on ${\cal S}$. The metric components have no dependence on $u$, and the timelike isometry $\partial/\partial u$ is null on the horizon at $r=0$. The black hole metric in a patch containing the horizon is given by
\bea
\label{gnulmet}
ds^2 = 2du dr + 2r h_i(r, y) du dy^i - r f(r, y) du^2 + ds_{\cal S}^2 \ .
\eea
The spatial horizon section ${\cal S}$ is given by $u=const,~ r=0$ with the metric
\bea
ds_{\cal S}^2  = \gamma_{i j}(r, y)dy^i dy^j \ .
\eea
We assume that ${\cal{S}}$ is compact, connected and without boundary. The
1-form $h$, scalar $\Delta$ and metric $\gamma$ are functions of $r$ and $y^{i}$; they are analytic in $r$ and regular at the horizon. The surface gravity associated with the Killing horizon is given by $\kappa = \frac{1}{2}f(y,0)$. The near-horizon limit is a particular decoupling limit defined by
\bea
\label{nhl}
r \rightarrow \epsilon r,~ u \rightarrow \epsilon^{-1} u,~ y^{i} \rightarrow y^{i}, \qquad {\rm and} \qquad \epsilon \rightarrow 0 \ .
\eea
This limit is only defined when $f(y,0) = 0$, which implies that the surface gravity vanishes, $\kappa = 0$. Hence the near horizon geometry is only well defined for extreme black holes,
and we shall consider only extremal black holes here. After taking the limit (\ref{nhl}) we obtain,
\bea
\label{nhm}
ds_{NH}^2 = 2du dr + 2r h_i(y) du dy^i - r^2 \Delta(y) du^2  + \gamma_{i j}(y)dy^i dy^j \ .
\eea
In particular, the form of the metric remains unchanged from (\ref{gnulmet}), however
the 1-form $h$, scalar $\Delta$ and metric $\gamma$ on ${\cal{S}}$ no longer have any radial dependence
{\footnote{The near-horizon metric (\ref{nhm}) also has a new scale symmetry, $r \rightarrow \lambda r,~ u \rightarrow \lambda^{-1}u$ generated by the Killing vector $L=u\partial_{u} - r\partial_{r}$. This, together with the Killing vector $V=\partial_u$ satisfy the algebra $[V, L] = V$ and they form a 2-dimensional non-abelian symmetry group ${\cal{G}}_2$. We shall show that this further enhances into a larger symmetry algebra, which will include a $\mathfrak{sl}(2,\mathbb{R})$ subalgebra.}}. For $N=(1,0)$, $D=6$ supergravity, in addition to the metric, there are also gauge field strengths and
scalars. We will assume that these are also analytic in $r$ and regular at the horizon,
and that there is also a consistent near-horizon limit for these matter fields:
\bea
A &=& -r \alpha \bbe^+ + {\tilde{A}}
\nonumber \\
F &=& \bbe^+ \wedge \bbe^- \alpha + r \bbe^+ \wedge T + {\tilde F}~,
\nonumber \\
H &=& \bbe^+ \wedge \bbe^- \wedge L + r \bbe^+ \wedge M + {\tilde H}
\la{hormetr}
\eea
where we have introduced the frame
\be
\label{basis1}
\bbe^+ = du, \qquad \bbe^- = dr + rh -\frac{1}{2} r^2 \Delta du, \qquad \bbe^i = e^i{}_j dy^j~,
\ee
in which the metric is
\bea
\label{nhf}
ds^2 &=&2 \bbe^+ \bbe^- + \delta_{ij} \bbe^i \bbe^j \ .
\eea

\subsection{Solving the KSEs along the Lightcone}

For a supersymmetric near-horizon geometry we assume there exists $\epsilon \neq 0$ solving the KSEs. We determine the necessary conditions on the Killing spinor by integrating along the two lightcone directions, i.e.\ along $u$ and $r$. To do this, we decompose $\epsilon$ as
\bea
\e=\e_++\e_-~,
\label{ksp1}
\eea
where $\Gamma_\pm\epsilon_\pm=0$, and find that
\bea\label{lightconesol}
\e_+=\phi_+(u,y)~,~~~\e_-=\phi_-+r \Gamma_-\Theta_+ \phi_+~,
\eea
and
\bea
\phi_-=\eta_-~,~~~\phi_+=\eta_++ u \Gamma_+ \Theta_-\eta_-~,
\eea
where
\bea
\label{thetapm}
\Theta_\pm &=& \tfrac{1}{4} h_i\Gamma^i \pm \frac{1}{8}e^{\frac{\Phi}{2}} L_{i}\Gamma^{i}
+ \frac{1}{48}e^{\frac{\Phi}{2}} {\tilde H}_{i j k}\Gamma^{i j k}
\eea
and $\eta_\pm$ depend only on the coordinates of the spatial horizon section ${\cal S}$.
Substituting the solution (\ref{lightconesol}) of the KSEs along the light cone directions back into the gravitino KSE (\ref{Gkse}), and appropriately expanding in the $r$ and $u$ coordinates, we find that
for the $\mu = \pm$ components, one obtains the additional conditions
\bea
\label{int1}
&&\bigg(\tfrac{1}{2}\Delta - \tfrac{1}{8}(dh)_{ij}\Gamma^{ij} + i g \alpha\bigg)\phi_+
\nonumber\\
&&+2\bigg(\tfrac{1}{4} h_i\Gamma^i - \tfrac{1}{8}e^{\frac{\Phi}{2}} L_{i}\Gamma^{i} + \tfrac{1}{48}e^{\frac{\Phi}{2}} {\tilde H}_{i j k}\Gamma^{i j k} \bigg)\tau_+ = 0~,
\eea
\bea
\label{int2}
\bigg(\frac{1}{4}\Delta h_i \Gamma^{i} - \frac{1}{4}\partial_{i}\Delta \Gamma^{i}\bigg)\phi_+ + \bigg(-\frac{1}{8}(dh)_{ij}\Gamma^{ij} +\frac{1}{8}e^{\frac{\Phi}{2}}M_{i j}\Gamma^{i j}\bigg) \tau_+ = 0~,
\eea
\bea
\label{int3}
\bigg(-\frac{1}{2}\Delta - \frac{1}{8}(dh)_{ij}\Gamma^{ij} + i g \alpha + \frac{1}{8}e^{\frac{\Phi}{2}}M_{i j}\Gamma^{i j}
-2\Theta_{+} \Theta_{-} \bigg)\phi_{-} = 0 \ .
\eea
Similarly the $\mu=i$ component of the gravitino KSEs gives
\bea
\label{int4}
\tilde{\nabla}_i \phi_\pm + \bigg( \mp \frac{1}{4}h_i  - i g {\tilde{A}}_{i} \mp \frac{1}{8}e^{\frac{\Phi}{2}} L_{j}\Gamma^{j}\Gamma_{i}
+ \frac{1}{48}e^{\frac{\Phi}{2}}{\tilde H}_{j k l}\Gamma^{j k l}\Gamma_{i}\bigg) \phi_\pm=0~,~~~
\eea
and
\bea
\label{int5}
&&\tilde \nabla_i \tau_{+} + \bigg( -\frac{3}{4}h_i - i g {\tilde A}_{i} + \frac{1}{8}e^{\frac{\Phi}{2}} L_{j}\Gamma^{j}\Gamma_{i} + \frac{1}{48}e^{\frac{\Phi}{2}} {\tilde H}_{j k l}\Gamma^{j k l}\Gamma_{i} \bigg)\tau_{+}
\nonumber \\
&&+ \bigg(-\frac{1}{4}(dh)_{ij}\Gamma^{j} + \frac{1}{16} e^{\frac{\Phi}{2}} M_{j k}\Gamma^{j k}\Gamma_{i}   \bigg)\phi_{+} = 0~,
\eea
where we have set
\bea
\label{int6}
\tau_{+} = \Theta_{+}\phi_{+} \ .
\eea
Similarly, substituting the solution of the KSEs (\ref{lightconesol}) into the algebraic KSE (\ref{Akse}) and expanding appropriately in the $u$ and $r$ coordinates, we find
\bea
\label{int7}
\bigg(\Gamma^{i}\nabla_{i}{\Phi} \pm e^{\frac{\Phi}{2}} L_{i}\Gamma^{i} - \frac{1}{6}e^{\frac{\Phi}{2}}{\tilde H}_{i j k}\Gamma^{i j k}\bigg)\phi_\pm = 0  \ ,
\eea
\be
\label{int8}
-\bigg(\Gamma^{i}\nabla_{i}{\Phi} - e^{\frac{\Phi}{2}} L_{i}\Gamma^{i} - \frac{1}{6}e^{\frac{\Phi}{2}}{\tilde H}_{i j k}\Gamma^{i j k} \bigg)\tau_{+} - \frac{1}{2}e^{\frac{\Phi}{2}}M_{i j}\Gamma^{i j} \phi_{+}=0~.
\ee
and (\ref{Fkse}),
\bea
\label{int9}
\bigg(e^{\frac{\Phi}{4}}(\mp 2  \alpha +
 {\tilde F}_{j k}\Gamma^{j k}) - 8 i g e^{-\frac{\Phi}{4}}\bigg) \phi_{\pm} = 0
\eea
\bea
\label{int10}
\bigg( e^{\frac{\Phi}{4}}(2\alpha + {\tilde F}_{j k}\Gamma^{j k}) - 8i g e^{-\frac{\Phi}{4}}\bigg)\tau_{+} + 2 e^{\frac{\Phi}{4}} T_{i}\Gamma^{i} \phi_+ = 0
\eea

In the following section we show that many of the above conditions are redundant: they are implied by the independent KSEs\footnote{These are the naive restrictions of the KSEs to ${\cal S}$.} (\ref{covr}) together with the field equations and Bianchi identities.

\newsection{Simplification of KSEs on ${\cal{S}}$}

The integrability conditions of the KSEs in any supergravity theory are known to imply some of the Bianchi identities and field equations. Also, the KSEs are first order differential equations which are usually easier to solve than the field equations which are second order. As a result, the standard approach to find solutions is to first solve all the KSEs and then impose the remaining independent components of the field equations and Bianchi identities as required.
We will take a different approach here because of the difficulty of solving the KSEs and the algebraic conditions which include the $\tau_+$ spinor given in (\ref{int6}). Furthermore, we are particularly interested
in the minimal set of conditions required for supersymmetry, in order to systematically analyse the necessary and
sufficient conditions for supersymmetry enhancement.

In particular, the conditions (\ref{int1}), (\ref{int2}), (\ref{int5}), and (\ref{int8}) which contain $\tau_+$ are implied from those containing $\phi_+$, along with some of the field equations and Bianchi identities. Furthermore, (\ref{int3}) and the terms linear in $u$ in (\ref{int4}), (\ref{int7}) and (\ref{int9}) from the $+$ component are implied by the field equations, Bianchi identities and the $-$ component of (\ref{int4}), (\ref{int7}) and (\ref{int9}).

A particular useful identity is obtained by considering the integrability condition of (\ref{int4}), which implies that
\bea
\label{DDphicond}
(\tilde{\nabla}_{j}\tilde{\nabla}_{i} - \tilde{\nabla}_{i}\tilde{\nabla}_{j})\phi_\pm &=& \bigg( \pm \tfrac{1}{4} \tilde{\nabla}_{j} (h_i) + i g {\tilde \nabla}_{j}(A_{i})
\nonumber\\&&
\pm \tfrac{1}{8}{\tilde \nabla}_{j}(e^{\frac{\Phi}{2}}L_{\ell})\Gamma^{\ell}\Gamma_{i}
- \tfrac{1}{48}{\tilde \nabla}_{j}(e^{\frac{\Phi}{2}} {\tilde H}_{\ell_1 \ell_2 \ell_3})\Gamma^{\ell_1 \ell_2 \ell_3}\Gamma_{i} \bigg) \phi_\pm
\nonumber \\
&+& \bigg(\pm \tfrac{1}{4}h_j  + i g {\tilde{A}}_{j} \pm \tfrac{1}{8}e^{\frac{\Phi}{2}} L_{\ell}\Gamma^{\ell}\Gamma_{j}
- \tfrac{1}{48}e^{\frac{\Phi}{2}}{\tilde H}_{\ell_1 \ell_2 \ell_3}\Gamma^{\ell_1 \ell_2 \ell_3}\Gamma_{j}\bigg)
\nonumber\\&&\times
\bigg(\pm \tfrac{1}{4}h_i  + i g {\tilde{A}}_{i} \pm \tfrac{1}{8}e^{\frac{\Phi}{2}} L_{k}\Gamma^{k}\Gamma_{i}
\nonumber \\
&&\quad- \tfrac{1}{48}e^{\frac{\Phi}{2}}{\tilde H}_{k_1 k_2 k_3}\Gamma^{k_1 k_2 k_3}\Gamma_{i}\bigg)\phi_\pm - (i \leftrightarrow j)
\eea
This will be used in the analysis of (\ref{int1}), (\ref{int3}), (\ref{int5}) and the positive chirality part of (\ref{int4}) which is linear in $u$. In order to show that the conditions are redundant, we will be considering different combinations of terms which vanish
as a consequence of the independent KSEs. However, non-trivial identities are found by
explicitly expanding out the terms in each case. Let us also define,
\bea
\label{int1condaux}
\mathcal{A}_1 = \bigg(\Gamma^{i}\nabla_{i}{\Phi} + e^{\frac{\Phi}{2}} L_{i}\Gamma^{i} - \frac{1}{6}e^{\frac{\Phi}{2}}{\tilde H}_{i j k}\Gamma^{i j k}\bigg)\phi_+  \ .
\eea
\bea
\label{int1condaux2}
\mathcal{B}_1 = \bigg(\Gamma^{i}\nabla_{i}{\Phi} - e^{\frac{\Phi}{2}} L_{i}\Gamma^{i} - \frac{1}{6}e^{\frac{\Phi}{2}}{\tilde H}_{i j k}\Gamma^{i j k}\bigg)\eta_- \ .
\eea

\bea
{\cal F}_1 = \bigg(e^{\frac{\Phi}{4}}(- 2  \alpha +
{\tilde F}_{j k}\Gamma^{j k}) - 8 i g e^{-\frac{\Phi}{4}}\bigg) \phi_{+}
\eea
\bea
{\cal G}_1 = \bigg(e^{\frac{\Phi}{4}}( 2  \alpha +
{\tilde F}_{j k}\Gamma^{j k}) - 8 i g e^{-\frac{\Phi}{4}}\bigg) \eta_{-}
\eea

\subsection{The condition (\ref{int1})}
\label{int1sec}
It can be shown that the algebraic condition on $\tau_+$ (\ref{int1}) is implied by the independent KSEs. Let us define,
\bea
\xi_1 &=& \bigg(\tfrac{1}{2}\Delta - \tfrac{1}{8}(dh)_{ij}\Gamma^{ij} + i g \alpha\bigg)\phi_+
\nonumber\\
&&+2\bigg(\tfrac{1}{4} h_i\Gamma^i - \tfrac{1}{8}e^{\frac{\Phi}{2}} L_{i}\Gamma^{i} + \tfrac{1}{48}e^{\frac{\Phi}{2}} H_{i j k}\Gamma^{i j k} \bigg)\tau_+\ ,
\eea
where $\xi_1=0$ is equal to the condition (\ref{int1}).
It is then possible to show that this expression for $\xi_1$ can be re-expressed as
\bea
\label{int1cond}
&&\xi_1 = \bigg(-\frac{1}{4}\tilde{R} - \Gamma^{i j}\tilde{\nabla}_{i}\tilde{\nabla}_{j}\bigg)\phi_+
+ \mu \mathcal{A}_1 + \lambda {\cal F}_1 = 0
\eea
where the first two terms cancel as a consequence of the definition of curvature, and
\bea
\mu &=& \frac{1}{16}{\tilde \nabla}_{i}{\Phi} \Gamma^{i} + \frac{1}{8}e^{\frac{\Phi}{2}} L_{i}\Gamma^{i} + \frac{1}{48}e^{\frac{\Phi}{2}}{\tilde H}_{i j k}\Gamma^{i j k}
\nonumber \\
\lambda &=& -\frac{3}{64}e^{\frac{\Phi}{4}}{\tilde F}_{i j}\Gamma^{i j} - \frac{5}{32}e^{\frac{\Phi}{4}}\alpha + \frac{1}{8}e^{-\frac{\Phi}{4}}g i
\eea
the scalar curvature can be written as
\bea
\tilde{R} &=& -2\Delta - \tfrac{1}{2}h^{2} + \tfrac{1}{4}{\tilde \nabla}^i{\Phi} {\tilde \nabla}_i{\Phi}
\nonumber\\
&+& \tfrac{5}{4}e^{\frac{\Phi}{2}}\alpha^2 + \tfrac{3}{8}e^{\frac{\Phi}{2}}\tilde{F}^2 + e^{\Phi}L^2 + \tfrac{1}{6}e^{\Phi}{\tilde H}^2 + 4e^{-\frac{\Phi}{2}}g^2 \ ,
\eea
The expression appearing in (\ref{int1condaux}) vanishes because
$\mathcal{A}_1 = {\cal F}_1 = 0$ is equivalent to the positive chirality part of (\ref{int7}) and (\ref{int9}).
Furthermore, the expression for $\xi_1$ given in (\ref{int1cond}) also vanishes.
We also use (\ref{DDphicond}) to evaluate the terms in the first bracket in (\ref{int1cond}) and explicitly expand out the terms with $\mathcal{A}_1$. In order to obtain (\ref{int1}) from these expressions we make use of the Bianchi identities (\ref{beq2}), the field equations (\ref{feq2}) and (\ref{feq3}). We have also made use of the $+-$ component of the Einstein equation (\ref{feq4}) in order to rewrite the scalar curvature $\tilde{R}$ in terms of $\Delta$. Therefore (\ref{int1}) follows from (\ref{int4}), (\ref{int7}) and (\ref{int9}) together with the field equations and Bianchi identities mentioned above.

\subsection{The condition (\ref{int2})}
The algebraic condition (\ref{int2}) follows from (\ref{int1}). It is convenient to define
\bea
\xi_2 = \bigg(\frac{1}{4}\Delta h_i \Gamma^{i} - \frac{1}{4}\partial_{i}\Delta \Gamma^{i}\bigg)\phi_+ + \bigg(-\frac{1}{8}(dh)_{ij}\Gamma^{ij} +\frac{1}{8}e^{\frac{\Phi}{2}}M_{i j}\Gamma^{i j}\bigg) \tau_+ \ ,
\eea
where $\xi_2=0$ equals the condition (\ref{int2}).
One can show after a computation that this expression for $\xi_2$ can be re-expressed as
\bea
\xi_2 = -\frac{1}{4}\Gamma^{i}\tilde{\nabla}_{i}{\xi_1} + \frac{7}{16}h_{j}\Gamma^{j}\xi_1 = 0 \ ,
\eea
which vanishes because $\xi_1 = 0$ is equivalent to the condition (\ref{int1}). In order to obtain this, we use the Dirac operator $\Gamma^{i}\tilde{\nabla}_{i}$ to act on (\ref{int1}) and apply the Bianchi identities (\ref{beq2}) with the field equations (\ref{feq2}) and (\ref{feq3}) to eliminate the terms which contain derivatives of the fluxes, and we can also use (\ref{int1}) to rewrite the $dh$-terms in terms of $\Delta$. We then impose the algebraic conditions (\ref{int7}) and (\ref{int8}) to eliminate the $\tilde{\nabla}_i \Phi$-terms, of which some of the remaining terms will vanish as a consequence of (\ref{int1}). We then obtain the condition (\ref{int2}) as required, therefore it follows from section \ref{int1sec} above that (\ref{int2}) is implied by (\ref{int4}) and (\ref{int7}) together with the field equations and Bianchi identities mentioned above.

\subsection{The condition (\ref{int5})}
The differential condition (\ref{int5}) is not independent. Let us define
\bea
\lambda_i &=& \tilde \nabla_i \tau_{+} + \bigg( -\frac{3}{4}h_i - i g {\tilde A}_{i} + \frac{1}{8}e^{\frac{\Phi}{2}} L_{j}\Gamma^{j}\Gamma_{i} + \frac{1}{48}e^{\frac{\Phi}{2}} {\tilde H}_{j k l}\Gamma^{j k l}\Gamma_{i} \bigg)\tau_{+}
\nonumber \\
&&+ \bigg(-\frac{1}{4}(dh)_{ij}\Gamma^{j} + \frac{1}{16} e^{\frac{\Phi}{2}} M_{j k}\Gamma^{j k}\Gamma_{i}   \bigg)\phi_{+} \ ,
\eea
where $\lambda_i=0$ is equivalent to the condition (\ref{int5}). We can re-express this expression for
$\lambda_i$ as
\bea
\label{int5cond}
\lambda_ i = \bigg(-\frac{1}{4}\tilde{R}_{i j}\Gamma^{j} + \frac{1}{2}\Gamma^{j}(\tilde{\nabla}_{j}\tilde{\nabla}_{i} - \tilde{\nabla}_{i}\tilde{\nabla}_{j}) \bigg)\phi_+  + \mu_{i}{\cal A}_{1} + \lambda_{i}{\cal F}_{1} = 0~,
\eea
where the first terms again cancel from the definition of curvature, and
\bea
\mu_i = \frac{1}{16}{\tilde \nabla}_{i}{\Phi} + \frac{1}{192}e^{\frac{\Phi}{2}}{\tilde H}_{\ell_1 \ell_2 \ell_3}\Gamma^{\ell_1 \ell_2 \ell_3}\Gamma_{i} - \frac{1}{32}e^{\frac{\Phi}{2}}L_\ell \Gamma^{\ell}\Gamma_{i}
\eea
and
\bea
\lambda_{i} = \frac{1}{128}e^{\frac{\Phi}{4}}{\tilde F}_{\ell_1 \ell_2}\Gamma^{\ell_1 \ell_2}\Gamma_{i} - \frac{1}{16}e^{\frac{\Phi}{4}}{\tilde F}_{i \ell}\Gamma^{\ell} - \frac{1}{64}e^{\frac{\Phi}{4}}\alpha \Gamma_{i} + \frac{1}{16}e^{-\frac{\Phi}{4}} g i \Gamma_{i}
\eea
This vanishes as $\mathcal{A}_1 = \mathcal{F}_1 = 0$ is equivalent to the positive chirality component of (\ref{int7}) and (\ref{int9}). The identity (\ref{int5cond}) is derived by making use of (\ref{DDphicond}), and explicitly expanding out the $\mathcal{A}_1$ and $\mathcal{F}_1$ terms. We can also evaluate (\ref{int5}) by substituting in (\ref{int6}) to eliminate $\tau_+$, and use (\ref{int4}) to evaluate the supercovariant derivative of $\phi_+$. Then, on adding this to (\ref{int5cond}), one obtains a condition which vanishes identically on making use of the Einstein equation (\ref{feq4}). Therefore it follows that (\ref{int5}) is implied by the positive chirality component of (\ref{int4}), (\ref{int6}) (\ref{int7}), the Bianchi identities (\ref{beq2}) and the gauge field equations (\ref{feq2}) and (\ref{feq3}).

\subsection{The condition (\ref{int8})}
The algebraic condition (\ref{int8}) follows from the independent KSEs. We define
\bea
\mathcal{A}_{2} &=& -\bigg(\Gamma^{i}\nabla_{i}{\Phi} - e^{\frac{\Phi}{2}} L_{i}\Gamma^{i} - \frac{1}{6}e^{\frac{\Phi}{2}}{\tilde H}_{i j k}\Gamma^{i j k} \bigg)\tau_{+} - \frac{1}{2}e^{\frac{\Phi}{2}}M_{i j}\Gamma^{i j} \phi_{+}
\eea
where $\mathcal{A}_{2}=0$ equals the expression in (\ref{int8}).
The expression for $\mathcal{A}_{2}$ can be rewritten as
\bea
\label{nnaux1a}
\mathcal{A}_{2} &=& -\frac{1}{2}\Gamma^{i}\tilde{\nabla}_{i}{({\cal A}_{1})}
+ \Phi_1 {\cal A}_{1} + \Phi_2 {\cal F}_{1}
\eea
where,
\bea
\Phi_{1} &=& \frac{3}{8}h_{\ell}\Gamma^{\ell} + \frac{i g}{2}{\cal A}_{\ell}  \Gamma^{\ell}
- \frac{1}{8}e^{\frac{\Phi}{2}}L_{\ell}\Gamma^{\ell}
+ \frac{1}{48}e^{\frac{\Phi}{2}}{\tilde H}_{\ell_1 \ell_2 \ell_3}\Gamma^{\ell_1 \ell_2 \ell_3}
\eea
and
\bea
\Phi_2 &=& -\frac{1}{16}e^{\frac{\Phi}{4}}{\tilde F}_{\ell_1 \ell_2}\Gamma^{\ell_1 \ell_2} + \frac{1}{8}\alpha e^{\frac{\Phi}{4}} - \frac{i g}{2}e^{-\frac{\Phi}{4}}
\eea
In evaluating the above conditions, we have made use of the $+$ component of (\ref{int4}) in order to evaluate the covariant derivative in the above expression. In addition we have made use of the Bianchi identities (\ref{beq2}) and the field equations (\ref{feq2}), (\ref{feq3}) and (\ref{feq6}).

It follows from (\ref{nnaux1a}) that $\mathcal{A}_{2}=0$ as a consequence of the condition $\mathcal{A}_{1} = \mathcal{F}_{1}=0$, which as we have already noted is equivalent to the positive chirality part of (\ref{int7}).

\subsection{The condition (\ref{int10})}
The algebraic condition (\ref{int10}) follows from the independent KSEs. We define
\bea
\mathcal{F}_{2} &=& \bigg( e^{\frac{\Phi}{4}}(2\alpha + {\tilde F}_{j k}\Gamma^{j k}) - 8i g e^{-\frac{\Phi}{4}}\bigg)\tau_{+} + 2 e^{\frac{\Phi}{4}} T_{i}\Gamma^{i} \phi_+
\eea
where $\mathcal{F}_{2}=0$ equals the expression in (\ref{int10}).
The expression for $\mathcal{F}_{2}$ can be rewritten as
\bea
\label{nnaux1b}
\mathcal{F}_{2} &=& -\frac{1}{2}\Gamma^{i}\tilde{\nabla}_{i}{({\cal F}_{1})}
+ \Phi_1 {\cal F}_{1} + \Phi_2 {\cal A}_{1}
\eea
where,
\bea
\Phi_{1} &=& \frac{3}{8}h_{\ell}\Gamma^{\ell} + \frac{i g}{2}{\cal A}_{\ell}  \Gamma^{\ell}
+ \frac{1}{8}e^{\frac{\Phi}{2}}L_{\ell}\Gamma^{\ell}
- \frac{1}{48}e^{\frac{\Phi}{2}}{\tilde H}_{\ell_1 \ell_2 \ell_3}\Gamma^{\ell_1 \ell_2 \ell_3}
\eea
and
\bea
\Phi_2 &=& \frac{1}{8}e^{\frac{\Phi}{4}}{\tilde F}_{\ell_1 \ell_2}\Gamma^{\ell_1 \ell_2} - \frac{1}{4}\alpha e^{\frac{\Phi}{4}} + i g e^{-\frac{\Phi}{4}}
\eea
In evaluating the above conditions, we have made use of the $+$ component of (\ref{int4}) in order to evaluate the covariant derivative in the above expression. In addition we have made use of the Bianchi identities (\ref{beq}) and the field equation (\ref{feq1}).

It follows from (\ref{nnaux1b}) that $\mathcal{F}_{2}=0$ as a consequence of the conditions $\mathcal{A}_{1} = \mathcal{F}_{1}=0$, which as we have already noted is equivalent to the positive chirality part of (\ref{int7}) and (\ref{int9}).

\subsection{The condition (\ref{int3})}
In order to show that (\ref{int3}) is implied by the independent KSEs, we define
\bea
\kappa &=& \bigg(-\frac{1}{2}\Delta - \frac{1}{8}(dh)_{ij}\Gamma^{ij} + i g \alpha + \frac{1}{8}e^{\frac{\Phi}{2}}M_{i j}\Gamma^{i j}
-2\Theta_{+} \Theta_{-} \bigg)\phi_{-} = 0  \ ,
\eea
where $\kappa$ equals the condition (\ref{int3}). Again, this expression can be rewritten as
\bea
\label{int3cond}
&&\xi_1 = \bigg(\frac{1}{4}\tilde{R} + \Gamma^{i j}\tilde{\nabla}_{i}\tilde{\nabla}_{j}\bigg)\phi_+
- \mu \mathcal{B}_1 - \lambda {\cal G}_1 = 0
\eea
where we use the (\ref{DDphicond}) to evaluate the terms in the first bracket, and
\bea
\mu &=& \frac{1}{16}{\tilde \nabla}_{i}{\Phi} \Gamma^{i} - \frac{1}{8}e^{\frac{\Phi}{2}} L_{i}\Gamma^{i} + \frac{1}{48}e^{\frac{\Phi}{2}}{\tilde H}_{i j k}\Gamma^{i j k}
\nonumber \\
\lambda &=& -\frac{3}{64}e^{\frac{\Phi}{4}}{\tilde F}_{i j}\Gamma^{i j} + \frac{5}{32}e^{\frac{\Phi}{4}}\alpha + \frac{1}{8}e^{-\frac{\Phi}{4}}g i
\eea
The expression above vanishes identically since the negative chirality component of (\ref{int7}) and (\ref{int9}) is equivalent to $\mathcal{B}_1 = \mathcal{G}_1 = 0$. In order to obtain (\ref{int3}) from these expressions we make use of the Bianchi identities (\ref{beq2}) and the field equations (\ref{feq3}),(\ref{feq4}) and (\ref{feq5}). Therefore (\ref{int3}) follows from (\ref{int4}), (\ref{int7}) and (\ref{int9}) together with the field equations and Bianchi identities mentioned above.

\subsection{The positive chirality part of (\ref{int4}) linear in $u$}
Since $\phi_+ = \eta_+ + u\Gamma_{+}\Theta_{-}\eta_-$, we must consider the part of the positive chirality component of (\ref{int4}) which is linear in $u$.
We then determine that $\mathcal{B}_{1}$ satisfies the following expression
\bea
\label{int4cond}
\bigg(\frac{1}{2}\Gamma^{j}(\tilde{\nabla}_{j}\tilde{\nabla}_{i} - \tilde{\nabla}_{i}\tilde{\nabla}_{j}) - \frac{1}{4}\tilde{R}_{i j}\Gamma^{j}  \bigg)\eta_{-} + \mu_i {\cal B}_1 + \lambda_i {\cal G}_1= 0 ~,
\eea
where
\bea
\mu_i = \frac{1}{16}{\tilde \nabla}_{i}{\Phi} + \frac{1}{192}e^{\frac{\Phi}{2}}{\tilde H}_{\ell_1 \ell_2 \ell_3}\Gamma^{\ell_1 \ell_2 \ell_3}\Gamma_{i} + \frac{1}{32}e^{\frac{\Phi}{2}}L_\ell \Gamma^{\ell}\Gamma_{i}
\eea
and
\bea
\lambda_{i} = \frac{1}{128}e^{\frac{\Phi}{4}}{\tilde F}_{\ell_1 \ell_2}\Gamma^{\ell_1 \ell_2}\Gamma_{i} - \frac{1}{16}e^{\frac{\Phi}{4}}{\tilde F}_{i \ell}\Gamma^{\ell} + \frac{1}{64}e^{\frac{\Phi}{4}}\alpha \Gamma_{i} + \frac{1}{16}e^{-\frac{\Phi}{4}} g i \Gamma_{i}
\eea
We note that $\mathcal{B}_{1}= \mathcal{G}_{1} = 0$ is equivalent to the negative chirality component of (\ref{int7}) and (\ref{int9}). Next, we use (\ref{DDphicond}) to evaluate the terms in the first bracket in (\ref{int4cond}) and explicitly expand out the terms with $\mathcal{B}_1$ and $\mathcal{G}_1$. The resulting expression corresponds to the expression obtained by expanding out the $u$-dependent part of the positive chirality component of (\ref{int4}) by using the negative chirality component of (\ref{int4}) to evaluate the covariant derivative. We have made use of the Bianchi identities (\ref{beq2}) and the gauge field equations (\ref{feq2}) and (\ref{feq3}).

\subsection{The positive chirality part of condition (\ref{int7}) linear in $u$}
Again, as $\phi_+ = \eta_+ + u\Gamma_{+}\Theta_{-}\eta_-$, we must consider the part of the positive chirality component of (\ref{int7}) which is linear in $u$.
One finds that the $u$-dependent part of (\ref{int7}) is proportional to
\bea
-\frac{1}{2}\Gamma^{i}\tilde{\nabla}_{i}{({\cal B}_{1})}
+ \Phi_1 {\cal B}_{1} + \Phi_2 {\cal G}_{1} \ ,
\eea
where,
\bea
\Phi_{1} &=& \frac{1}{8}h_{\ell}\Gamma^{\ell} + \frac{i g}{2}{\cal A}_{\ell}  \Gamma^{\ell}
+ \frac{1}{8}e^{\frac{\Phi}{2}}L_{\ell}\Gamma^{\ell}
+ \frac{1}{48}e^{\frac{\Phi}{2}}{\tilde H}_{\ell_1 \ell_2 \ell_3}\Gamma^{\ell_1 \ell_2 \ell_3}
\eea
and
\bea
\Phi_2 &=& -\frac{1}{16}e^{\frac{\Phi}{4}}{\tilde F}_{\ell_1 \ell_2}\Gamma^{\ell_1 \ell_2} - \frac{1}{8}\alpha e^{\frac{\Phi}{4}} - \frac{i g}{2}e^{-\frac{\Phi}{4}}
\eea
and where we use the (\ref{DDphicond}) to evaluate the terms in the first bracket. In addition we have made use of the Bianchi identities (\ref{beq2}) and the field equations (\ref{feq2}), (\ref{feq3}) and (\ref{feq6}).

\subsection{The positive chirality part of condition (\ref{int9}) linear in $u$}

Finally, we must consider the part of the positive chirality component of (\ref{int9}) which is linear in $u$.
One finds that the $u$-dependent part of (\ref{int9}) is proportional to
\bea
-\frac{1}{2}\Gamma^{i}\tilde{\nabla}_{i}{({\cal F}_{1})}
+ \Phi_1 {\cal B}_{1} + \Phi_2 {\cal G}_{1}
\eea
where,
\bea
\Phi_{1} &=& \frac{1}{8}h_{\ell}\Gamma^{\ell} + \frac{i g}{2}{\cal A}_{\ell}  \Gamma^{\ell}
- \frac{1}{8}e^{\frac{\Phi}{2}}L_{\ell}\Gamma^{\ell}
- \frac{1}{48}e^{\frac{\Phi}{2}}{\tilde H}_{\ell_1 \ell_2 \ell_3}\Gamma^{\ell_1 \ell_2 \ell_3}
\eea
and
\bea
\Phi_2 &=& \frac{1}{8}e^{\frac{\Phi}{4}}{\tilde F}_{\ell_1 \ell_2}\Gamma^{\ell_1 \ell_2} + \frac{1}{4}\alpha e^{\frac{\Phi}{4}} + i g e^{-\frac{\Phi}{4}}
\eea
In evaluating the above conditions, we have made use of the $+$ component of (\ref{int4}) in order to evaluate the covariant derivative in the above expression. In addition we have made use of the Bianchi identities (\ref{beq}) and the field equation (\ref{feq1}).

\subsection{The Independent KSEs on $\cal{S}$}

On taking the previous sections into account, it follows that, on making use of the field equations and Bianchi identities, the independent KSEs are
\bea
\label{covr}
\nabla^{(\pm)}_{i} \eta_{\pm} = 0, \qquad {\cal A}^{(\pm)}\eta_{\pm} = 0 \qquad {\cal F}^{(\pm)}\eta_{\pm} = 0
\eea
where
\bea
\nabla^{(\pm)}_{i} = \tilde{\nabla}_{i} + \Psi^{(\pm)}_{i}
\eea
with
\bea
\label{alg1pm}
\Psi^{(\pm)}_{i} &=& \mp \frac{1}{4}h_i  - i g {\tilde{A}}_{i} \mp \frac{1}{8}e^{\frac{\Phi}{2}} L_{j}\Gamma^{j}\Gamma_{i}
+ \frac{1}{48}e^{\frac{\Phi}{2}}{\tilde H}_{j k l}\Gamma^{j k l}\Gamma_{i} \ ,
\eea
and
\bea
\label{alg2pm}
\mathcal{A}^{(\pm)} &=& \Gamma^{i}\nabla_{i}{\Phi} \pm e^{\frac{\Phi}{2}} L_{i}\Gamma^{i} - \frac{1}{6}e^{\frac{\Phi}{2}}{\tilde H}_{i j k}\Gamma^{i j k} \ ,
\eea
\bea
\label{alg3pm}
{\cal F}^{(\pm)} &=& e^{\frac{\Phi}{4}}(\mp 2  \alpha +
{\tilde F}_{j k}\Gamma^{j k}) - 8 i g e^{-\frac{\Phi}{4}}
\eea
These are derived from the naive restriction of the supercovariant derivative and the algebraic KSE on ${\cal S}$.
Furthermore, if $\eta_{-}$ solves (\ref{covr}) then
\bea
\eta_+ = \Gamma_{+}\Theta_{-}\eta_{-}~,
\label{epfem}
\eea
also solves (\ref{covr}). However, further analysis using global techniques, is required in order to determine if $\Theta_-$
has a non-trivial kernel.

\newsection{Global Analysis: Lichnerowicz Theorems}
\label{maxpex}

In this section, we shall establish a correspondence between parallel spinors $\eta_\pm$ satisfying
(\ref{covr}), and spinors in the kernel of appropriately defined horizon Dirac operators.
We define the horizon Dirac operators associated with the supercovariant derivatives following from the gravitino KSE as
\bea
{\mathscr D}^{(\pm)} \equiv \Gamma^{i}\nabla_{i}^{(\pm)} = \Gamma^{i}\tilde{\nabla}_{i} + \Psi^{(\pm)}~,
\eea
where
\bea
\label{Psipm}
\Psi^{(\pm)} \equiv \Gamma^{i}\Psi^{(\pm)}_{i} = \mp \frac{1}{4}h_i\Gamma^{i}  - i g{\tilde A}_{i}\Gamma^{i} \pm \frac{1}{4}e^{\frac{\Phi}{2}}L_{i}\Gamma^{i} + \frac{1}{24}e^{\frac{\Phi}{2}}{\tilde H}_{i j k}\Gamma^{i j k} \ .
\eea

To establish the Lichnerowicz-type theorems, we begin by computing the Laplacian of $\parallel \eta_\pm \parallel^2$. Here we will assume throughout that ${\mathscr D}^{(\pm)}\eta_\pm=0$, so
\bea
\label{lapla}
\tilde{\nabla}^i \tilde{\nabla}_i ||\eta_{\pm}||^2 = 2{\rm Re } \langle\eta_\pm,\tilde{\nabla}^i \tilde{\nabla}_i\eta_\pm\rangle + 2 {\rm Re } \langle\tilde{\nabla}^i \eta_\pm, \tilde{\nabla}_i \eta_\pm\rangle \ .
\eea
To evaluate this expression note that
\bea
\tilde{\nabla}^i \tilde{\nabla}_i \eta_\pm &=& \Gamma^{i}\tilde{\nabla}_{i}(\Gamma^{j}\tilde{\nabla}_j \eta_\pm) -\Gamma^{i j}\tilde{\nabla}_i \tilde{\nabla}_j \eta_\pm
\nonumber \\
&=& \Gamma^{i}\tilde{\nabla}_{i}(\Gamma^{j}\tilde{\nabla}_j \eta_\pm) + \frac{1}{4}\tilde{R}\eta_\pm
\nonumber \\
&=& \Gamma^{i}\tilde{\nabla}_{i}(-\Psi^{(\pm)}\eta_\pm) + \frac{1}{4}\tilde{R} \eta_\pm \ .
\eea
Therefore the first term in (\ref{lapla}) can be written as,
\bea
\label{lap1}
{\rm Re } \langle\eta_\pm,\tilde{\nabla}^i \tilde{\nabla}_i\eta_\pm \rangle &=& \frac{1}{4}\tilde{R}\parallel \eta_\pm \parallel^2
+ {\rm Re } \langle\eta_\pm, \Gamma^{i}\tilde{\nabla}_i(-\Psi^{(\pm)})\eta_\pm\rangle
\nonumber\\
&&+ {\rm Re } \langle\eta_\pm, \Gamma^{i}(-\Psi^{(\pm)})\tilde{\nabla}_i \eta_\pm \rangle~.
\eea
For the second term in (\ref{lapla}) we write,
\bea
\label{lap2}
\hspace{-1cm}{\rm Re } \langle\tilde{\nabla}^i \eta_\pm, \tilde{\nabla}_i \eta_\pm\rangle
&=& \parallel {\nabla^{(\pm)}}\eta_{\pm} \parallel^2 - 2{\rm Re } \langle \eta_{\pm}, \Psi^{(\pm)i\dagger}\tilde{\nabla}_{i}\eta_{\pm}\rangle - {\rm Re } \langle \eta_\pm , \Psi^{(\pm)i\dagger}\Psi^{(\pm)}_i \eta_\pm \rangle  .
\eea
We remark that $\dagger$ is the adjoint with respect to the $Spin_{c}(4)$-invariant inner product ${\rm Re } \langle \phantom{i},\phantom{i} \rangle$.\footnote{This inner product is positive definite and symmetric.}
Therefore using (\ref{lap1}) and (\ref{lap2}) with (\ref{lapla}) we have,
\bea
\label{extralap1b}
\frac{1}{2}\tilde{\nabla}^i \tilde{\nabla}_i ||\eta_{\pm}||^2 &=& \parallel {\nabla^{(\pm)}}\eta_{\pm} \parallel^2 + {\rm Re } \langle \eta_\pm, \bigg(\tfrac{1}{4}\tilde{R} + \Gamma^{i}\tilde{\nabla}_i(-\Psi^{(\pm)})
\nonumber\\&&\quad
- \Psi^{(\pm)i\dagger}\Psi^{(\pm)}_i \bigg) \eta_\pm \rangle
\nonumber \\
&+& {\rm Re } \langle \eta_\pm, \bigg( \Gamma^{i}(-\Psi^{(\pm)}) - 2\Psi^{(\pm)i\dagger}\bigg)\tilde{\nabla}_i \eta_\pm \rangle \ .
\eea
In order to simplify the expression for the Laplacian, we observe that the second line in (\ref{extralap1b}) can be rewritten as
\bea
\label{bilin}
{\rm Re } \langle \eta_\pm, \bigg( \Gamma^{i}(-\Psi^{(\pm)}) - 2\Psi^{(\pm)i\dagger}\bigg)\tilde{\nabla}_i \eta_\pm \rangle &=& {\rm Re } \langle \eta_\pm, \mathcal{K}^{(\pm)}\Gamma^{i}\tilde{\nabla}_i \eta_\pm \rangle
\nonumber\\
&&\pm\tfrac{1}{2}h^i \tilde{\nabla}_i \parallel \eta_\pm \parallel^2~,
\eea
where
\bea
\mathcal{K}^{(\pm)} = \mp\frac{1}{4}h_{j}\Gamma^{j} - i g {\tilde A}_{i}\Gamma^{i}
\eea
We also have the following identities
\bea
\label{hermiden1}
{\rm Re } \langle \eta_+, \Gamma^{\ell_1 \ell_2} \eta_+ \rangle = {\rm Re } \langle \eta_+, \Gamma^{\ell_1 \ell_2 \ell_3} \eta_+ \rangle = 0
\eea
and
\bea
\label{hermiden2}
{\rm Re } \langle \eta_+, i\Gamma^{\ell} \eta_+ \rangle = 0 \  .
\eea
It follows that
\bea
\label{laplacian}
\frac{1}{2}\tilde{\nabla}^i \tilde{\nabla}_i \parallel \eta_\pm \parallel^2 &=& \parallel {{\nabla}^{(\pm)}}\eta_{\pm} \parallel^2 \pm\frac{1}{2}h^i \tilde{\nabla}_{i}\parallel \eta_\pm \parallel^2
\nonumber \\
&+& {\rm Re } \langle \eta_\pm, \bigg(\frac{1}{4}\tilde{R} + \Gamma^{i}\tilde{\nabla}_i(-\Psi^{(\pm)})
\nonumber\\&&\quad
- \Psi^{(\pm)i\dagger}\Psi^{(\pm)}_i  + \mathcal{K}^{(\pm)}(-\Psi^{(\pm)})\bigg) \eta_\pm \rangle \ ,
\eea
It is also useful to evaluate ${\tilde{R}}$ using (\ref{feq4}); we obtain
\bea
\tilde{R} &=& -\tilde{\nabla}^{i}(h_i) + \tfrac{1}{2}h^2 + \tfrac{1}{4}{\tilde \nabla}^{i}{\Phi}{\tilde \nabla}_{i}{\Phi}
\nonumber\\
&+& \tfrac{1}{4}e^{\frac{\Phi}{2}}{\tilde F}^2 + \tfrac{1}{2}e^{\frac{\Phi}{2}}\alpha^2 + \tfrac{1}{12}e^{\Phi}{\tilde H}^2 + \tfrac{1}{2}e^{\Phi}L^2 + 8 e^{-\frac{\Phi}{2}}g^2,
\eea
One obtains, upon using the field equations and Bianchi identities,
\bea
\label{quad}
&&\bigg(\frac{1}{4}\tilde{R} + \Gamma^{i}\tilde{\nabla}_i(-\Psi^{(\pm)})
\nonumber\\&&\quad
- \Psi^{(\pm)i\dagger}\Psi^{(\pm)}_i  + \mathcal{K}^{(\pm)}(-\Psi^{(\pm)}) \bigg)\eta_\pm
\nonumber \\
&=&\bigg[ i {\tilde \nabla}^{i}{{\tilde A}_{i}} \pm \frac{i g}{4}e^{\frac{\Phi}{2}}{\tilde A}^{i}L_{i}
\mp \frac{i}{2} g {\tilde A}^{i}h_{i}
\nonumber\\&&
+(\pm \frac{1}{4}{\tilde \nabla}_{\ell_1}{h_{\ell_2}} - \frac{1}{16} e^{\frac{\Phi}{2}} L_{\ell_1}h_{\ell_2}
- \frac{1}{8} e^{\frac{\Phi}{2}} {\tilde \nabla}^{i}{H_{\ell_1 \ell_2 i}}
\nonumber \\
&\mp& \frac{1}{4}e^{\frac{\Phi}{2}}{\tilde \nabla}_{\ell_1}L_{\ell_2} - \frac{1}{16}e^{\frac{\Phi}{2}}{\tilde H}_{\ell_1 \ell_2 k}{\tilde \nabla}^{k}{\Phi} \pm \frac{1}{32}e^{\frac{\Phi}{2}}{\tilde H}_{\ell_1 \ell_2 k}h^{k}
\pm \frac{1}{8} e^{\frac{\Phi}{2}} L_{\ell_1}{\tilde \nabla}_{\ell_2}{\Phi})\Gamma^{\ell_1 \ell_2}
\nonumber \\
&+& \frac{i g}{24} e^{\frac{\Phi}{2}} {\tilde A}_{\ell_1}H_{\ell_2 \ell_3 \ell_4}\Gamma^{\ell_1 \ell_2 \ell_3 \ell_4}\bigg]\eta_\pm
\nonumber \\
&+& \bigg(\frac{1}{16}{\tilde \nabla}^{i}{\Phi}{\tilde \nabla}_{i}{\Phi}
\pm \frac{1}{8}e^{\frac{\Phi}{2}} L^{i}{\tilde \nabla}_{i}{\Phi}
\nonumber\\&&
+ \frac{1}{48}e^{\frac{\Phi}{2}}{\tilde H}_{\ell_1 \ell_2 \ell_3}{\tilde \nabla}_{\ell_4}{\Phi}\Gamma^{\ell_1 \ell_2 \ell_3 \ell_4} + \frac{1}{16} e^{\Phi} L^2
\nonumber \\
&\pm& \frac{1}{48}e^{\Phi}{\tilde H}_{\ell_1 \ell_2 \ell_3}L_{\ell_4}\Gamma^{\ell_1 \ell_2 \ell_3 \ell_4}
\nonumber\\&&
- \frac{1}{64}e^{\Phi} {\tilde H}_{i \ell_1 \ell_2} {\tilde H}^{i}{}_{\ell_3 \ell_4}\Gamma^{\ell_1 \ell_2 \ell_3 \ell_4}
+ \frac{1}{96} e^{\Phi} {\tilde H}^2\bigg)\eta_\pm
\nonumber \\
&+& \bigg(\frac{1}{8}e^{\frac{\Phi}{2}} \alpha^2 - \frac{1}{32}e^{\frac{\Phi}{2}} {\tilde F}_{\ell_1 \ell_2}{\tilde F}_{\ell_3 \ell_4}\Gamma^{\ell_1 \ell_2 \ell_3 \ell_4}
\nonumber\\&&
+ \frac{1}{16}e^{\frac{\Phi}{2}}{\tilde F}^2 + \frac{i g}{2}{\tilde F}_{\ell_1 \ell_2}\Gamma^{\ell_1 \ell_2} + 2 e^{-\frac{\Phi}{2}}g^2\bigg)\eta_\pm
\nonumber \\
&-& \frac{1}{4} \big(1 \mp 1\big) {\tilde{\nabla}}^i (h_i)  \eta_\pm  \ .
\eea
One can show that the fourth and fifth line in (\ref{quad}) can be written in terms of the algebraic KSE (\ref{alg2pm}), in particular we find,
\bea
\label{blah}
\frac{1}{16}{\cal A}^{(\pm)\dagger}{\cal A}^{(\pm)}\eta_\pm &=& \frac{1}{16}{\tilde \nabla}^{i}{\Phi}{\tilde \nabla}_{i}{\Phi}
\pm \frac{1}{8}e^{\frac{\Phi}{2}} L^{i}{\tilde \nabla}_{i}{\Phi}
\nonumber\\&&
+ \frac{1}{48}e^{\frac{\Phi}{2}}{\tilde H}_{\ell_1 \ell_2 \ell_3}{\tilde \nabla}_{\ell_4}{\Phi}\Gamma^{\ell_1 \ell_2 \ell_3 \ell_4} + \frac{1}{16} e^{\Phi} L^2
\nonumber \\
&\pm& \frac{1}{48}e^{\Phi}{\tilde H}_{\ell_1 \ell_2 \ell_3}L_{\ell_4}\Gamma^{\ell_1 \ell_2 \ell_3 \ell_4}
\nonumber\\&&
- \frac{1}{64}e^{\Phi} {\tilde H}_{i \ell_1 \ell_2} {\tilde H}^{i}{}_{\ell_3 \ell_4}\Gamma^{\ell_1 \ell_2 \ell_3 \ell_4}
+ \frac{1}{96} e^{\Phi} {\tilde H}^2
\eea
and the sixth line,
\bea
\frac{1}{32}{\cal F}^{(\pm)\dagger}{\cal F}^{(\pm)}\eta_\pm &=& \frac{1}{8}e^{\frac{\Phi}{2}} \alpha^2 - \frac{1}{32}e^{\frac{\Phi}{2}} {\tilde F}_{\ell_1 \ell_2}{\tilde F}_{\ell_3 \ell_4}\Gamma^{\ell_1 \ell_2 \ell_3 \ell_4}
\nonumber \\
&&+ \frac{1}{16}e^{\frac{\Phi}{2}}{\tilde F}^2 + \frac{i g}{2}{\tilde F}_{\ell_1 \ell_2}\Gamma^{\ell_1 \ell_2} + 2 e^{-\frac{\Phi}{2}}g^2
\eea
Note that on using (\ref{hermiden1}) and (\ref{hermiden2}) all the terms on the RHS of the above expression, with the exception of the final four lines, vanish in the second line of (\ref{laplacian}) since all these terms in (\ref{quad}) are anti-Hermitian.
Also, for $\eta_+$ the final line in (\ref{quad}) also vanishes and thus there is no contribution to the Laplacian of $\parallel \eta_+ \parallel^2$ in (\ref{laplacian}). For $\eta_{-}$ the final line in (\ref{quad}) does give an extra term in the Laplacian of $\parallel \eta_- \parallel^2$ in (\ref{laplacian}). For this reason, the analysis of the conditions imposed by the global properties of ${\cal{S}}$ is different in these two cases and thus we will consider the Laplacians of $\parallel \eta_\pm \parallel^2$ separately.

{\sloppy
\begin{theorem}[Lichnerowicz theorem for $\eta_+$]
Let $\mathcal{S}$ be compact, connected and without boundary, and let $\eta_+$ satisfy $\mathscr{D}^{(+)}\eta_+=0$. Then $\eta_+$ is a Killing spinor on $\mathcal{S}$, i.e., $\nabla^{(+)}\eta_+=0$,
$\mathcal{A}^{(+)}\eta_+=0$, $\mathcal{F}^{(+)}\eta_+=0$, and $\|\eta_+\|=\mathrm{const}$.
\end{theorem}
}

\noindent{\it Proof.}
For the Laplacian of $\parallel \eta_+ \parallel^2$, we obtain from (\ref{laplacian}):
\bea
\label{l1}
{\tilde{\nabla}}^{i}{\tilde{\nabla}}_{i}\parallel\eta_+\parallel^2 - h^i {\tilde{\nabla}}_{i}\parallel\eta_+\parallel^2 &=& 2\parallel{{\nabla}^{(+)}}\eta_{+}\parallel^2
\nonumber\\
&+& \tfrac{1}{8}\parallel{\cal A}^{(+)}\eta_+ \parallel^2 + \tfrac{1}{16}\parallel{\cal F}^{(+)}\eta_+ \parallel^2
\eea
The maximum principle thus implies that $\eta_+$ are Killing spinors on ${\cal{S}}$ assuming that it is compact, connected and without boundary, i.e.
\bea
{{\nabla}^{(+)}}\eta_{+}=0, \quad {\cal A}^{(+)}\eta_{+} = 0, \quad {\cal F}^{(+)}\eta_{+} = 0
\eea
and moreover $\parallel\eta_+\parallel=\mathrm{const}$. \hfill$\square$

{\sloppy
\begin{theorem}[Lichnerowicz theorem for $\eta_-$]
Let $\mathcal{S}$ be compact, connected and without boundary, and let $\eta_-$ satisfy $\mathscr{D}^{(-)}\eta_-=0$. Then $\eta_-$ is a Killing spinor on $\mathcal{S}$, i.e., $\nabla^{(-)}\eta_-=0$,
$\mathcal{A}^{(-)}\eta_-=0$, $\mathcal{F}^{(-)}\eta_-=0$.
\end{theorem}
}

\noindent{\it Proof.}
The Laplacian of $\parallel \eta_- \parallel^2$
is calculated from (\ref{laplacian}), on taking account of the contribution to the second line of
(\ref{laplacian}) from the final line of (\ref{quad}). One
obtains
\bea
\label{l2}
{\tilde{\nabla}}^{i} (W_{i})
= 2\parallel{{\nabla}^{(-)}}\eta_{-}\parallel^2 +~ \frac{1}{8}\parallel{\cal A}^{(-)}\eta_- \parallel^2 +~ \frac{1}{16}\parallel{\cal F}^{(-)}\eta_- \parallel^2
\eea
where $W = d\parallel \eta_- \parallel^2 + \parallel \eta_- \parallel^2  h$. On integrating this over ${\cal{S}}$ and assuming that ${\cal{S}}$ is compact and without boundary, the LHS vanishes since it is a total derivative and one finds that $\eta_{-}$ are Killing spinors on ${\cal{S}}$, i.e.
\bea
{{\nabla}^{(-)}}\eta_{-}=0, \quad {\cal A}^{(-)}\eta_{-} = 0, \quad {\cal F}^{(-)}\eta_{-} = 0
\eea
\hfill$\square$

This establishes the Lichnerowicz type theorems for both positive and negative chirality spinors $\eta_\pm$ which are in the kernels of the horizon Dirac operators ${{\mathscr D}}^{(\pm)}$: i.e.

\bea
\{ \ {{\nabla}^{(\pm)}}\eta_{\pm}=0, \quad {\cal A}^{(\pm)}\eta_{\pm} = 0, \quad {\rm and} \quad {\cal F}^{(\pm)}\eta_{\pm} = 0 \ \}
\quad \Longleftrightarrow \quad {{\mathscr D}}^{(\pm)} \eta_\pm = 0 \ .
\eea

\newsection{(Super)symmetry Enhancement}

We now turn to the counting of supersymmetries. Let $N_\pm$ denote the number of linearly
independent $\eta_\pm$ Killing spinors, equivalently
\bea
N_\pm = \dim \Ker\big\{\nabla^{(\pm)}, {\cal A}^{(\pm)}, {\cal F}^{(\pm)}\big\}~.
\eea
For the ungauged theory the horizon spinors take values in the $Spin(4)$ bundles $\mathbb S^\pm$,
while in the gauged theory they take values in the $Spin_c(4)$ bundles
$\mathbb S^\pm\otimes \mathcal L$, where $\mathcal L$ is the $U(1)$ line bundle determined by the
horizon gauge field. By the Lichnerowicz theorems of section~5,
\bea
N_\pm = \dim\Ker\,\mathscr D^{(\pm)}~.
\eea

{\sloppy
\begin{proposition}[Supersymmetry counting]
Assume that the near-horizon data are smooth, that $\mathcal S$ is compact, connected and without
boundary, and that the horizon field equations and Bianchi identities hold. Then the total number
of supersymmetries is
\bea
N = 2N_- + \mathrm{Index}(\mathscr D^{(+)})~.
\eea
\end{proposition}
}

\noindent{\it Proof.}
Since $\mathscr D^{(+)}$ is defined on the even-dimensional manifold $\mathcal S$,
\bea
\mathrm{Index}(\mathscr D^{(+)}) = \dim\Ker\,\mathscr D^{(+)} - \dim\Ker\,(\mathscr D^{(+)})^\dagger~.
\eea
Moreover,
\bea
\Gamma_- (\mathscr D^{(+)})^\dagger = \mathscr D^{(-)}\Gamma_-~,
\eea
so $\dim\Ker\,(\mathscr D^{(+)})^\dagger = \dim\Ker\,\mathscr D^{(-)} = N_-$. Using also
$N_+=\dim\Ker\,\mathscr D^{(+)}$, one obtains
\bea
\mathrm{Index}(\mathscr D^{(+)}) = N_+ - N_-~,
\eea
and hence $N=N_++N_-=2N_-+\mathrm{Index}(\mathscr D^{(+)})$. \hfill$\square$

\begin{remark}
This proposition is unconditional for the class of regular horizons considered here. The only later
conditional statement in the paper concerns the gauged $\mathfrak{sl}(2,\mathbb R)$ enhancement,
which depends on the additional assumption $\mathrm{Ker}\,\Theta_- = \{0\}$.
\end{remark}

\subsection{The index contribution to supersymmetry counting}
\label{sec:index}

A central result established above is that the number of supersymmetries preserved by a smooth
compact horizon section $\mathcal{S}$ is
\bea
N = 2N_- + \mathrm{Index}\!\left(\mathscr{D}^{(+)}\right),
\label{eq:index_counting}
\eea
where $\mathscr{D}^{(+)}$ is the horizon Dirac operator associated to the positive lightcone
chirality sector. More precisely, as stated in the introduction, this is the index of a Dirac
operator twisted by a vector bundle $E$ over $\mathcal{S}$, whose precise form depends on the
gauge structure of the supergravity theory under consideration.

In the present $D=6$ theory the proof of the supersymmetry-counting formula only requires the
abstract index $\mathrm{Index}(\mathscr{D}^{(+)})$, and does not require an explicit topological
evaluation. In particular, the generalized Lichnerowicz-type theorem established in section~5
implies that the zero modes of $\mathscr{D}^{(+)}$ are in one-to-one correspondence with the
relevant positive-chirality Killing spinors on $\mathcal{S}$, and hence the contribution of this
sector is measured by the index of $\mathscr{D}^{(+)}$.

This is to be contrasted with type IIA supergravity \cite{iiaindex}, where $\mathscr{D}^{(+)}$
acts on the Majorana non-Weyl spinor bundle and maps $S_+$ to $S_+$ (same sector), so its
principal symbol coincides with that of the Dirac operator on Majorana spinors and the index
vanishes. Similarly, for $D=11$ M-theory the spatial horizon section is 9-dimensional and the
index vanishes for any Dirac operator on an odd-dimensional manifold \cite{11index, atiyah1}.
In the present $D=6$ chiral theory, neither obstruction applies.

In the ungauged theory, where the twisting is trivial, $\mathscr{D}^{(+)}$ has the same principal
symbol as the ordinary chiral Dirac operator on the spin bundle $\mathbb{S}^{+}$ over the compact
four-manifold $\mathcal{S}$. In that case, the Atiyah--Singer index theorem \cite{atiyah1} gives
\bea
\label{indexformula}
\mathrm{Index}\!\left(\mathscr{D}^{(+)}\right)
=
\int_{\mathcal{S}} \hat{A}(T\mathcal{S})
=
-\frac{\mathrm{sign}(\mathcal{S})}{8}~,
\label{eq:ungauged_dirac_index}
\eea
where we have used the Hirzebruch signature theorem
\bea
\int_{\mathcal{S}} p_1(T\mathcal{S}) = 3\,\mathrm{sign}(\mathcal{S})
\eea
together with the degree-4 expansion $\hat{A}(T\mathcal{S}) = -p_1(T\mathcal{S})/24 + \cdots$.
Note that this index is an integer for any compact oriented $Spin(4)$ manifold $\mathcal{S}$. In
fact, since $\mathcal{S}$ admits a spin structure (as required for the horizon spinors $\eta_\pm$
to exist), Rokhlin's theorem implies $\mathrm{sign}(\mathcal{S}) \in 16\mathbb{Z}$, so
$\mathrm{Index}(\mathscr{D}^{(+)}) = -\mathrm{sign}(\mathcal{S})/8 \in 2\mathbb{Z}$ in the
ungauged theory.

For example, if $\mathcal{S}=K3$, then $\mathrm{sign}(K3)=-16$ and
\bea
\mathrm{Index}\!\left(\mathscr{D}^{(+)}\right)=2~,
\eea
consistent with $K3$ admitting exactly 2 parallel Weyl spinors. For $\mathcal{S}=T^4$ all
characteristic classes vanish and
\bea
\mathrm{Index}\!\left(\mathscr{D}^{(+)}\right)=0~,
\eea
reproducing $N=2N_-$.

In the gauged theory, an explicit evaluation of $\mathrm{Index}(\mathscr{D}^{(+)})$ requires the
precise identification of the twisting bundle $E$ induced by the $U(1)$ connection appearing in
the horizon supercovariant derivative, together with its charge normalization. Since the proof of
(\ref{eq:index_counting}) does not depend on such an explicit identification, we shall leave the
gauged index in abstract form.

If, in a given class of examples, $\mathscr{D}^{(+)}$ is identified with a spin Dirac operator
twisted by a complex line bundle $\mathcal{L}$, then the Atiyah--Singer theorem yields
\bea
\mathrm{Index}\!\left(\mathscr{D}^{(+)}\right)
=
\int_{\mathcal{S}} \hat{A}(T\mathcal{S})\,\mathrm{ch}(\mathcal{L})
=
-\frac{\mathrm{sign}(\mathcal{S})}{8}
+\frac{1}{2}\,c_1(\mathcal{L})^2[\mathcal{S}]~,
\label{eq:twisted_dirac_index_conditional}
\eea
but we shall not assume such an identification in the general gauged case.

\begin{remark}
\label{rem:parity_gauged}
Even without an explicit identification of $\mathcal{L}$, one can draw a parity conclusion. Since
$\mathcal{S}$ is a spin 4-manifold, its intersection form is even: for every
$x\in H^2(\mathcal{S};\mathbb{Z})$ one has $x^2[\mathcal{S}]\in 2\mathbb{Z}$. Applying this to
$x = c_1(\mathcal{L})$ gives $c_1(\mathcal{L})^2[\mathcal{S}]\in 2\mathbb{Z}$, so
$\frac{1}{2}c_1(\mathcal{L})^2[\mathcal{S}]\in\mathbb{Z}$. This shows the second term in
(\ref{eq:twisted_dirac_index_conditional}) is an integer, but not necessarily even. However,
since Rokhlin's theorem gives $-\mathrm{sign}(\mathcal{S})/8\in 2\mathbb{Z}$, the parity of the
full index is controlled by the second term alone:
\bea
\mathrm{Index}\!\left(\mathscr{D}^{(+)}\right) \equiv \tfrac{1}{2}\,c_1(\mathcal{L})^2[\mathcal{S}]
\pmod{2}~.
\eea
In particular, $\mathrm{Index}(\mathscr{D}^{(+)})$ is even whenever $c_1(\mathcal{L})\in
2H^2(\mathcal{S};\mathbb{Z})$: if $c_1(\mathcal{L})=2y$ for some $y\in H^2(\mathcal{S};\mathbb{Z})$,
then $c_1(\mathcal{L})^2[\mathcal{S}] = 4y^2[\mathcal{S}]\in 8\mathbb{Z}$ (using the evenness of
the intersection form again), so $\frac{1}{2}c_1(\mathcal{L})^2[\mathcal{S}]\in 4\mathbb{Z}$ and
the index lies in $2\mathbb{Z}$.
\end{remark}

\subsection{Algebraic Relationship between $\eta_+$ and $\eta_{-}$ Spinors}
\label{kernal}

The map $\eta_-\mapsto \eta_+ = \Gamma_+\Theta_-\eta_-$ is the mechanism which relates
negative- and positive-lightcone chirality Killing spinors. It is therefore central both to the
supersymmetry-counting formula above and to the symmetry-enhancement statement below. The key
question is whether $\Theta_-$ can have a non-trivial kernel.

{\sloppy
\begin{proposition}[Ungauged triviality of $\mathrm{Ker}\,\Theta_-$]
\label{prop:ker}
Assume $g=0$. Suppose $\mathrm{Ker}\,\Theta_-\neq\{0\}$. Then all horizon fluxes vanish, the dilaton is
constant, and the near-horizon data are trivial. In particular, the resulting spacetime geometry is
locally $\mathbb R^{1,1}\times T^4$.
\end{proposition}
}

\noindent{\it Proof.}
Suppose that there exists $\eta_-\neq 0$ such that $\Theta_-\eta_-=0$. Then (\ref{int3}) gives
$\Delta\,\mathrm{Re}\langle\eta_-,\eta_-\rangle=0$, so $\Delta=0$ because $\eta_-$ is nowhere
vanishing. The gravitino KSE $\nabla^{(-)}\eta_-=0$, together with
$\mathrm{Re}\langle\eta_-,\Gamma_i\Theta_-\eta_-\rangle=0$, implies that
\begin{eqnarray}
\label{nrm1a}
{\tilde{\nabla}}_i \parallel \eta_-\parallel^2 = - h_i  \parallel \eta_-\parallel^2~.
\end{eqnarray}
Hence $dh=0$, and then (\ref{feq7}) implies that $T=M=0$. Taking the divergence of (\ref{nrm1a}),
eliminating ${\tilde{\nabla}}^i h_i$ via (\ref{feq3}), and using $\Delta=0$, one finds
\begin{eqnarray}
\label{nrm1ab}
{\tilde{\nabla}}^i {\tilde{\nabla}}_i  \parallel \eta_-\parallel^2 &=&
 \bigg(\tfrac{3}{8}e^{\frac{\Phi}{2}} \alpha^2 + \tfrac{1}{16}e^{\frac{\Phi}{2}}\tilde{F}^2  + \tfrac{1}{4}e^{\Phi}L^2
 \nonumber\\&&
 + \tfrac{1}{12}e^{\Phi}{\tilde H}^2 - 2e^{-\frac{\Phi}{2}} g^2 \bigg)  \parallel \eta_-\parallel^2~.
\end{eqnarray}
For the ungauged theory, $g=0$, so the maximum principle implies that $\parallel\eta_-\parallel^2$
is constant. Thus $\alpha=\tilde F=L=\tilde H=0$, and then (\ref{int7}) implies that $\Phi$ is
constant. Finally, integrating (\ref{feq3}) over $\mathcal S$ gives $h=0$. Hence all fluxes vanish,
the scalar is constant, and the near-horizon geometry is locally $\mathbb R^{1,1}\times T^4$.
\hfill$\square$

\begin{remark}
For the gauged theory the same argument does \emph{not} go through, because the final term in
(\ref{nrm1ab}) has negative sign and obstructs the maximum principle. Therefore the triviality of
$\mathrm{Ker}\,\Theta_-$ is proved only in the ungauged case. Whenever we discuss symmetry
enhancement in the gauged theory below, $\mathrm{Ker}\,\Theta_- = \{0\}$ is an additional
hypothesis rather than a theorem.
\end{remark}

\subsection{The $\mathfrak{sl}(2,\bR)$ Symmetry}

\begin{remark}
In this subsection we assume $N_-\neq0$ and use the paired Killing spinor
$\eta_+=\Gamma_+\Theta_-\eta_-$. For ungauged horizons with non-trivial fluxes this is automatic by
proposition~\ref{prop:ker}; for gauged horizons it requires the additional hypothesis
$\mathrm{Ker}\,\Theta_- = \{0\}$.
\end{remark}

Having established how to obtain $\eta_+$ type spinors from $\eta_-$ spinors, we next proceed
to determine the $\mathfrak{sl}(2,\bR)$ spacetime symmetry.
First note that the spacetime Killing spinor $\epsilon$ can be expressed in terms of $\eta_\pm$ as
\begin{eqnarray}
\epsilon= \eta_++ u \Gamma_+\Theta_-\eta_-+ \eta_-+r \Gamma_-\Theta_+\eta_++ru \Gamma_-\Theta_+\Gamma_+\Theta_-\eta_-~.
\label{gensolkse}
\end{eqnarray}
Since the $\eta_-$ and $\eta_+$ Killing spinors appear in pairs for supersymmetric horizons, let us choose a $\eta_-$ Killing spinor. Then from the previous results, horizons with non-trivial fluxes also admit $\eta_+=\Gamma_+\Theta_-\eta_-$ as a Killing spinor. Taking $\eta_-$ and $\eta_+=\Gamma_+\Theta_-\eta_-$,
one can construct two linearly independent Killing spinors on the spacetime as
\bea
\epsilon_1=\eta_-+u\eta_++ru \Gamma_-\Theta_+\eta_+~,~~~\epsilon_2=\eta_++r\Gamma_-\Theta_+\eta_+~.
\eea
It is known from the general theory of supersymmetric $D=6$ backgrounds that for any Killing spinors $\zeta_1$ and $\zeta_2$ the dual vector field $K(\zeta_1, \zeta_2)$ of the 1-form
bilinear
\bea
\omega(\zeta_1, \zeta_2) &=& {\rm Re } \langle(\Gamma_+-\Gamma_-) \zeta_1, \Gamma_a\zeta_2\rangle\, e^a
\label{1formbi}
\eea
is a Killing vector which leaves invariant all the other bosonic fields of the theory. Evaluating the 1-form bilinears of the Killing spinor $\epsilon_1$ and $\epsilon_2$, we find that
\begin{eqnarray}
\label{bforms}
\hspace{-1cm}\omega_1(\epsilon_1, \epsilon_2)&=& (2r {\rm Re} \langle\Gamma_+\eta_-, \Theta_+\eta_+\rangle+  4 u r^2  \parallel \Theta_{+}\eta_+\parallel^2) \,{\bf{e}}^+-2u \parallel\eta_+\parallel^2\, {\bf{e}}^-
\cr
&+& ({\rm Re} \langle \Gamma_+\eta_{-}, \Gamma_{i}\eta_{+}\rangle + 4 u r {\rm Re} \langle \eta_+, \Gamma_{i} \Theta_+ \eta_+ \rangle) {\bf{e}}^i~,
 \cr
 \omega_2(\epsilon_2, \epsilon_2)&=& 4 r^2 \parallel \Theta_+ \eta_+\parallel^2 \,{\bf{e}}^+-2 \parallel\eta_+\parallel^2 {\bf{e}}^- + 4 r{\rm Re}  \langle \eta_{+}, \Gamma_{i} \Theta_+ \eta_+ \rangle {\bf{e}}^i~,
 \cr
 \omega_3(\epsilon_1, \epsilon_1)&=&(2\parallel\eta_-\parallel^2+4r u {\rm Re} \langle\Gamma_+\eta_-, \Theta_+\eta_+\rangle+ 4 r^2 u^2 \parallel \Theta_+ \eta_+\parallel^2 ) {\bf{e}}^+
 \cr
 &-& 2u^2 \parallel\eta_+\parallel^2 {\bf{e}}^-+(2u {\rm Re } \langle \Gamma_+ \eta_- , \Gamma_i \eta_+ \rangle
+ 4 u^2 r{\rm Re} \langle \eta_+, \Gamma_i \Theta_+ \eta_+ \rangle) {\bf{e}}^i~.
\nonumber \\
\end{eqnarray}
We can establish the following identities
\begin{eqnarray}
\label{ident1}
- \Delta\, \parallel\eta_+\parallel^2 +4  \parallel\Theta_+ \eta_+\parallel^2 =0~,~~~{\rm Re } \langle \eta_+ , \Gamma_i \Theta_+ \eta_+ \rangle  =0~,
\end{eqnarray}
which follow from the first integrability condition in (\ref{int1}), $\parallel\eta_+\parallel=\mathrm{const}$ and the KSEs of $\eta_+$. Further simplification to the bilinears can be obtained by making use of (\ref{ident1}).
We then obtain
\begin{eqnarray}
 \omega_1(\epsilon_1, \epsilon_2)&=& (2r {\rm Re} \langle\Gamma_+\eta_-, \Theta_+\eta_+\rangle+ u r^2 \Delta \parallel \eta_+\parallel^2) \,{\bf{e}}^+
 \cr
 &&-2u \parallel\eta_+\parallel^2\, {\bf{e}}^-+ \tilde V_i {\bf{e}}^i~,
 \cr
 \omega_2(\epsilon_2, \epsilon_2)&=& r^2 \Delta\parallel\eta_+\parallel^2 \,{\bf{e}}^+-2 \parallel\eta_+\parallel^2 {\bf{e}}^-~,
 \cr
 \omega_3(\epsilon_1, \epsilon_1)&=&(2\parallel\eta_-\parallel^2+4r u {\rm Re} \langle\Gamma_+\eta_-, \Theta_+\eta_+\rangle+ r^2 u^2 \Delta \parallel\eta_+\parallel^2) {\bf{e}}^+
 \cr
 && \qquad\qquad \qquad\qquad -2u^2 \parallel\eta_+\parallel^2 {\bf{e}}^-+2u \tilde V_i {\bf{e}}^i~,
 \label{b1forms}
\end{eqnarray}
where we have set
\begin{eqnarray}
\label{vii}
\tilde V_i =  {\rm Re } \langle \Gamma_+ \eta_- , \Gamma_i \eta_+ \rangle\, ~.
\end{eqnarray}

To uncover explicitly the $\mathfrak{sl}(2,\mathbb{R})$ symmetry of such horizons it remains to compute the Lie bracket algebra of the vector fields $K_1$, $K_2$ and $K_3$ which are dual to the 1-form spinor bilinears $\omega_1, \omega_2$ and $\omega_3$. In simplifying the resulting expressions, we shall make use of the following identities
\begin{eqnarray}
&&-2 \parallel\eta_+\parallel^2-h_i \tilde V^i+2 {\rm Re } \langle\Gamma_+\eta_-, \Theta_+\eta_+\rangle=0~,
\cr
&&i_{\tilde V} (dh)+2 d {\rm Re } \langle\Gamma_+\eta_-, \Theta_+\eta_+\rangle=0~,
\cr
&& 2 {\rm Re } \langle\Gamma_+\eta_-, \Theta_+\eta_+\rangle-\Delta \parallel\eta_-\parallel^2=0~,
\cr
&&{\tilde V}+ \parallel\eta_-\parallel^2 h+d \parallel\eta_-\parallel^2=0~.
\label{conconx}
\end{eqnarray}

We then obtain the following dual Killing vector fields:
\begin{eqnarray}
K_1 &=&-2u \parallel\eta_+\parallel^2 \partial_u+ 2r \parallel\eta_+\parallel^2 \partial_r+ \tilde V~,
\cr
K_2 &=&-2 \parallel\eta_+\parallel^2 \partial_u~,
\cr
K_3 &=&-2u^2 \parallel\eta_+\parallel^2 \partial_u +(2 \parallel\eta_-\parallel^2+ 4ru \parallel\eta_+\parallel^2)\partial_r+ 2u \tilde V~.
\label{kkk}
\end{eqnarray}

As we have previously mentioned, each of these Killing vectors also leaves invariant all the other
bosonic fields in the theory. It is then straightforward to determine the algebra satisfied by
these isometries:

\begin{theorem}[Bracket algebra]
The Lie bracket algebra of $K_1$, $K_2$ and $K_3$ is $\mathfrak{sl}(2,\bR)$.
\end{theorem}

\noindent{\it Proof.} Using the identities summarised above, one can demonstrate after a direct computation that
\bea
{}[K_1,K_2]&=&2 \parallel\eta_+\parallel^2 K_2~,
\nonumber \\
{}[K_2, K_3]&=&-4 \parallel\eta_+\parallel^2 K_1~,
\nonumber \\
{}[K_3,K_1]&=&2 \parallel\eta_+\parallel^2 K_3~.
\eea
\hfill$\square$

{\sloppy
\begin{corollary}[Ungauged symmetry enhancement]
Assume $g=0$, that the horizon fluxes are non-trivial, and that $N_-\neq 0$. Then the isometry
algebra of the near-horizon spacetime contains an $\mathfrak{sl}(2,\mathbb R)$ subalgebra.
\end{corollary}
}

{\sloppy
\begin{corollary}[Conditional gauged symmetry enhancement]
Assume $g\neq0$, that the horizon fluxes are non-trivial, that $N_-\neq0$, and that
$\mathrm{Ker}\,\Theta_- = \{0\}$. Then the isometry algebra of the near-horizon spacetime contains
an $\mathfrak{sl}(2,\mathbb R)$ subalgebra.
\end{corollary}
}

A special case arises for $\tilde V=0$, where the group action generated by $K_1, K_2$ and $K_3$ has only 2-dimensional orbits. A direct substitution of this condition in (\ref{conconx}) reveals that
\bea
\label{zva}
\Delta \parallel\eta_-\parallel^2=2 \parallel\eta_+\parallel^2~,~~~h=\Delta^{-1} d\Delta~.
\eea
Since $h$ is exact, such horizons are static. A coordinate transformation $r\rightarrow \Delta r$ reveals that the geometry is a warped product of $AdS_2$ with ${\cal S}$, $AdS_2\times_w {\cal S}$.

\subsection{Isometries of ${\cal S}$}

It is known that the vector fields associated with the 1-form Killing spinor bilinears given in (\ref{1formbi}) leave invariant all the fields of
gauged $D=6$ supergravity. In particular suppose that $\tilde V \neq 0$. The isometries $K_a$ ($a=1,2,3$) leave all the bosonic fields invariant:
\bea
{\cal L}_{K_a} g=0, \qquad {\cal L}_{K_a} F=0, \qquad {\cal L}_{K_a} H=0, \qquad {\cal L}_{K_a} \Phi=0 \ .
\eea
Imposing these conditions and expanding in $u,r$, and also making use of the identities
(\ref{conconx}), one finds that
\begin{eqnarray}
&&\tilde\nabla_{(i} \tilde V_{j)}=0~,\quad {\cal L}_{\tilde V} h= {\cal L}_{\tilde V}\Delta=0~,\quad {\cal L}_{\tilde V} \Phi = 0~,
\nonumber\\
&&{\cal L}_{\tilde V} \tilde{F}= {\cal L}_{\tilde V} \alpha= {\cal L}_{\tilde V} L= {\cal L}_{\tilde V} \tilde{H}=0~.
\nonumber \\
\end{eqnarray}
Therefore $\tilde V$ is an isometry of ${\cal S}$ and leaves all the fluxes on ${\cal S}$ invariant. In fact, ${\tilde{V}}$ is a spacetime
isometry as well. Furthermore, the conditions (\ref{conconx}) imply that ${\cal L}_{\tilde V}\parallel\eta_-\parallel^2=0$.

\subsection{Conditions on the geometry}

We consider the further restrictions on the geometry of ${\cal S}$. We begin by explicitly expanding out the identities established in (\ref{ident1}), which follow from the first integrability condition in (\ref{int1}), $\parallel\eta_+\parallel=\mathrm{const}$ and the KSEs of $\eta_+$, in terms of bosonic fields and using (\ref{conconx}) along with the field equations (\ref{feq1})--(\ref{feq6}) and Bianchi identities (\ref{beq}) and (\ref{beq2}). On expanding (\ref{ident1}) we obtain,
\bea
\label{bos1}
\Delta\, \parallel\eta_+\parallel^2 &=& {\rm Re } \langle \eta_+ , \bigg(\tfrac{1}{4}h^2 + \tfrac{1}{4}e^{\frac{\Phi}{2}}h_{i}L^i + \tfrac{1}{16}e^{\Phi}L^2 + \tfrac{1}{96}e^{\Phi}H^2
\nonumber \\
&&+(- \tfrac{1}{24}e^{\frac{\Phi}{2}}{\tilde H}_{\ell_1 \ell_2 \ell_3 \ell_4}h_{\ell_4}
\nonumber \\
&-& \tfrac{1}{48}e^{\Phi}{\tilde H}_{\ell_1 \ell_2 \ell_3 \ell_4}L_{\ell_4} - \tfrac{1}{64}e^{\Phi}{\tilde H}^{k}{}_{\ell_1 \ell_2}{\tilde H}_{k \ell_3 \ell_4})\Gamma^{\ell_1 \ell_2 \ell_3 \ell_4}
\bigg) \eta_+ \rangle \ ,
\eea
and
\bea
\label{bos2}
{\rm Re } \langle \eta_+ , \Gamma_i \Theta_+ \eta_+ \rangle = {\rm Re } \langle \eta_+ , \bigg(\frac{1}{4}h_{i}  + \frac{1}{8}e^{\frac{\Phi}{2}} L_i + \frac{1}{48} e^{\frac{\Phi}{2}} {\tilde H}_{\ell_1 \ell_2 \ell_3}\Gamma_{i}{}^{\ell_1 \ell_2 \ell_3}\bigg) \eta_+ \rangle = 0 \ .
\eea
On contracting and substituting this in (\ref{bos1}) we can write,
\bea
\Delta\, \parallel\eta_+\parallel^2 &=& {\rm Re } \langle \eta_+, \bigg(-\frac{1}{4}h^2 -\frac{1}{4}e^{\frac{\Phi}{2}} L^i h_i - \frac{1}{16}e^{\Phi}L^2 + \frac{1}{96}e^{\Phi}{\tilde H}^2
\nonumber \\
&-& \frac{1}{64}e^{\Phi}{\tilde H}^{k}{}_{\ell_1 \ell_2}{\tilde H}_{k \ell_3 \ell_4}\Gamma^{\ell_1 \ell_2 \ell_3 \ell_4}\bigg)\eta_+ \rangle
\eea
From the algebraic KSE (\ref{alg2pm}) we have,
\bea
{\rm Re } \langle \eta_\pm , {\cal A}^{(\pm)}\eta_\pm \rangle &=&({\tilde \nabla}_i \Phi \pm e^{\frac{\Phi}{2}}L_i){\rm Re } \langle \eta_\pm, \Gamma^{i} \eta_{\pm} \rangle = 0
\nonumber \\
{\rm Re } \langle \eta_\pm , \Gamma_{i}{\cal A}^{(\pm)}\eta_\pm \rangle &=& {\rm Re } \langle \eta_\pm, \bigg( {\tilde \nabla}_i \Phi \pm e^{\frac{\Phi}{2}}L_i - \frac{1}{6}e^{\frac{\Phi}{2}}{\tilde H}_{\ell_1 \ell_2 \ell_3}\Gamma_{i}{}^{\ell_1 \ell_2 \ell_3} \bigg) \eta_{\pm} \rangle = 0
\eea
From this and (\ref{bos2}) we obtain,
\bea
{\rm Re } \langle \eta_+ , \bigg(\Gamma_i \Theta_+ + \frac{1}{8}\Gamma_{i}{\cal A}^{(+)}\bigg) \eta_+ \rangle = \bigg(\frac{1}{4}h_i + \frac{1}{4}e^{\frac{\Phi}{2}} L_i + \frac{1}{8}{\tilde \nabla}_{i}{\Phi}\bigg)\parallel\eta_+\parallel^2 = 0
\eea
since $\eta_+ \neq 0$ the norm is non-vanishing and we can write,
\bea
\label{expr1}
h_{i} = -\bigg(e^{\frac{\Phi}{2}}L_i + \frac{1}{2}{\tilde \nabla}_i {\Phi}\bigg)
\eea
On taking the divergence of this expression and using the field equations (\ref{feq2}), (\ref{feq4}) and (\ref{feq6}) and substituting back (\ref{expr1}), we obtain the condition,
\bea
\label{del}
\Delta = \frac{1}{2}e^{\frac{\Phi}{2}}\alpha^2
\eea
On considering the algebraic KSE (\ref{alg3pm}) we have,
\bea
{\rm Re } \langle \eta_\pm , {\cal F}^{(\pm)}\eta_\pm \rangle &=& \mp 2 e^{\frac{\Phi}{4}} \alpha \parallel\eta_\pm \parallel^2 = 0
\nonumber \\
{\rm Re } \langle \eta_\pm , \Gamma_{i}{\cal F}^{(\pm)}\eta_\pm \rangle &=& 2e^{\frac{\Phi}{4}}  {\tilde F}_{i \ell}{\rm Re } \langle \eta_\pm, \Gamma^{\ell} \eta_{\pm} \rangle = 0
\eea
Thus we obtain $\alpha = 0$ and from (\ref{del}) this implies $\Delta = 0$ which from (\ref{conconx}) implies ${\rm Re } \langle\Gamma_+\eta_-, \Theta_+\eta_+\rangle = 0$. The other identities in (\ref{conconx}) become,
\begin{eqnarray}
\hspace{-1cm}&&-2 \parallel\eta_+\parallel^2-h_i \tilde V^i=0~,~~~i_{\tilde V} (dh)=0~,~~~ {\tilde V}+ \parallel\eta_-\parallel^2 h+d \parallel\eta_-\parallel^2=0~.
\end{eqnarray}
Using these identities it is straightforward to show that there are no near-horizon geometries for which $h=0$ or ${\tilde V} = 0$ since this would lead to a contradiction to our assumption that $\eta_{+} \neq 0$.

\newsection{Conclusion}

We have analysed supersymmetric near-horizon geometries of $N=(1,0)$, $D=6$ gauged and ungauged
supergravity by solving the KSEs along the lightcone directions, reducing the independent horizon
system to a set of equations on the compact spatial section $\mathcal S$, and establishing
Lichnerowicz-type theorems for both horizon Dirac operators $\mathscr D^{(\pm)}$. The strongest
unconditional result is the supersymmetry-counting theorem
\begin{eqnarray}
N = 2N_- + \mathrm{Index}(\mathscr D^{(+)})~,
\end{eqnarray}
valid for smooth horizons with compact, connected, boundaryless $\mathcal S$ satisfying the horizon
field equations and Bianchi identities.

A key feature of the six-dimensional theory is that the relevant index need not vanish. Because
$\mathcal S$ is four-dimensional and the theory is chiral, the horizon Dirac operator is genuinely
chiral. In the ungauged theory this gives
$\mathrm{Index}(\mathscr D^{(+)})=-\mathrm{sign}(\mathcal S)/8$ explicitly via the
Atiyah--Singer theorem. Since $\mathcal{S}$ is spin, Rokhlin's theorem forces
$\mathrm{sign}(\mathcal{S}) \in 16\mathbb{Z}$, so the index is always an even integer $-2k$ with
$k = \mathrm{sign}(\mathcal{S})/16\in\mathbb{Z}$, and the total supersymmetry count
$N = 2(N_- - k)$ is manifestly even. In the gauged theory the index receives additional contributions from the $U(1)$ gauge sector
and is left in abstract form. If the twisting bundle can be identified with a complex line bundle
$\mathcal{L}$, the even intersection form on the spin manifold $\mathcal{S}$ forces
$c_1(\mathcal{L})^2[\mathcal{S}]\in 2\mathbb{Z}$, so the index is an integer. Its parity is
controlled by $\frac{1}{2}c_1(\mathcal{L})^2[\mathcal{S}]$, and the index is even whenever
$c_1(\mathcal{L})\in 2H^2(\mathcal{S};\mathbb{Z})$; see Remark~\ref{rem:parity_gauged}.
This distinguishes the present analysis from the earlier $D=11$ and type-IIA cases, where the
index vanishes.

The symmetry-enhancement statement requires a more careful formulation. In the ungauged theory, if
the fluxes are non-trivial and $N_-\neq0$, then $\mathrm{Ker}\,\Theta_- = \{0\}$ follows from a
maximum-principle argument, and the near-horizon spacetime admits an
$\mathfrak{sl}(2,\mathbb R)$ symmetry algebra. In the gauged theory the same conclusion is obtained
only under the additional hypothesis $\mathrm{Ker}\,\Theta_- = \{0\}$; the negative gauging term in
(\ref{nrm1ab}) prevents us from promoting this hypothesis to a theorem by the methods used here.
Accordingly, we do not claim an unconditional proof of the full gauged horizon conjecture.

There are several natural directions for further work. The most immediate is to determine whether
$\mathrm{Ker}\,\Theta_- = \{0\}$ can be proved directly in the gauged theory, thereby completing the
symmetry-enhancement argument without additional hypotheses. It would also be worthwhile to extend
the analysis to $(1,0)$ theories with more general matter couplings, in particular additional tensor,
vector, and hypermultiplet sectors, and to compare the resulting global constraints with the local
classification results already available in the literature \cite{6d16,6d23}.

\setcounter{section}{0}\setcounter{equation}{0}
\appendix{Supersymmetry Conventions}

We follow the spinor conventions of \cite{6d1, 6d2} with mostly positive signature. The $8\times 8$ Dirac matrices in six dimensions obey the Clifford algebra,
\bea
\{ \Gamma_{M}, \Gamma_{N} \} = 2g_{M N}
\eea
The chirality projector is defined as,
\bea
\Gamma_{*} = \Gamma_0 \cdots \Gamma_5, ~~~ \Gamma_{*}^2 = 1,~~~ \Gamma_{*}^{\dagger}=-\Gamma_{*}
\eea
The gamma matrices also satisfy the duality relation,
\bea
\Gamma^{A_1 \cdots A_n} = \frac{(-1)^{[n/2]}}{(6-n)!}\epsilon^{A_1 \cdots A_n B_1 \cdots B_{6-n}}\Gamma_{B_1 \cdots B_{6-n}} \Gamma_{*}
\eea
with $\epsilon^{012345}=1$. For a product of two anti-symmetrized gamma matrices we have,
\bea
\Gamma_{A_1 \cdots A_n}\Gamma^{B_1 \cdots B_m} = \sum_{k=0}^{min(n,m)}\frac{m! n!}{(m-k)! (n-k)! k!}\Gamma_{[A_1 \cdots A_{n-k}}^{\phantom{i} [B_{k+1}\cdots B_{m}}\delta^{B_1\cdots}_{A_n}\delta^{B_k]}_{A_{n-k+1}]} \ .
\eea
All the spinors are symplectic Majorana,
\bea
\chi^\alpha = \epsilon^{\alpha \beta}(\bar \chi)^T_\beta,~~ {\bar \chi}_\alpha = (\chi^\alpha)^{\dagger}\Gamma_0
\eea
where ${\bar \chi}^\alpha = ({\chi^\alpha})^T$ and $\alpha, \beta$ are $Sp(1)$ indices. It will be convenient to decompose the spinors into positive and negative chiralities
with respect to the lightcone directions as
\bea
\epsilon = \epsilon_+ + \epsilon_-~,
\eea
where
\bea
\Gamma_{+-} \epsilon_\pm = \pm \epsilon_\pm \ , \qquad {\rm or \ equivalently} \qquad \Gamma_\pm \epsilon_\pm =0~.
\eea
The representation of $Spin(5,1)$ decomposes under $Spin(4)=SU(2) \times SU(2)$ specified by the lightcone projections $\Gamma_\pm$. We have also made use of the $Spin(4)$-invariant inner product ${\rm Re } \langle , \rangle$ which is identified with the standard Hermitian inner product. In particular, note that $(\Gamma_{ij})^\dagger = - \Gamma_{ij}$.

\appendix{Spin Connection and Curvature}

The non-vanishing components of the spin connection in the frame basis (\ref{basis1}) are
\begin{eqnarray}
&&\Omega_{-,+i} = -\tfrac{1}{2} h_i~,~~~
\Omega_{+,+-} = -r \Delta, \quad \Omega_{+,+i} =\tfrac{1}{2} r^2(  \Delta h_i - \partial_i \Delta),
\cr
&&\Omega_{+,-i} = -\tfrac{1}{2} h_i, \quad \Omega_{+,ij} = -\tfrac{1}{2} r dh_{ij}~,~~~
\Omega_{i,+-} = \tfrac{1}{2} h_i, \quad \Omega_{i,+j} = -\tfrac{1}{2} r dh_{ij},
\cr
&&\Omega_{i,jk}= \tilde\Omega_{i,jk}~,
\end{eqnarray}
where $\tilde\Omega$ denotes the spin-connection of the 4-manifold ${\cal{S}}$ with basis ${\bf{e}}^i$.
If $f$ is any function of spacetime, then frame derivatives are expressed in terms of co-ordinate derivatives as
\begin{eqnarray}
\label{frco}
\partial_+ f &=& \partial_u f +\tfrac{1}{2} r^2 \Delta \partial_r f~,~~
\partial_- f = \partial_r f~,~~
\partial_i f = {\tilde{\partial}}_i f -r \partial_r f h_i \ .
\end{eqnarray}
The non-vanishing components of the Ricci tensor in the
 basis (\ref{basis1}) are
\bea
R_{+-} &=& \tfrac{1}{2} \tn^i h_i - \Delta -\tfrac{1}{2} h^2~,~~~
R_{ij} = {\tilde{R}}_{ij} + \tn_{(i} h_{j)} -\tfrac{1}{2} h_i h_j
\nonumber \\
R_{++} &=& r^2 \bigg( \tfrac{1}{2} \tn^2 \Delta -\tfrac{3}{2} h^i \tn_i \Delta -\tfrac{1}{2} \Delta \tn^i h_i + \Delta h^2
+\tfrac{1}{4} (dh)_{ij} (dh)^{ij} \bigg)
\nonumber \\
R_{+i} &=& r \bigg( \tfrac{1}{2} \tn^j (dh)_{ij} - (dh)_{ij} h^j - \tn_i \Delta + \Delta h_i \bigg) \ ,
\eea
where $\tn$ denotes the Levi-Civita connection of ${\cal S}$, ${\tilde{R}}$ is the Ricci tensor of the horizon section ${\cal S}$, and $i,j$ denote $\bbe^i$ frame indices.

\appendix{Horizon Bianchi Identities and Field Equations}
Substituting the fields (\ref{nhf}) into the Bianchi identity $dF = 0$ and $dH = \frac{1}{2}F \wedge F$ implies
\bea
\label{beq}
T = (d_h \alpha), ~~d\tilde{F} = 0
\eea
and
\bea
\label{beq2}
M = (d_h L) - \alpha {\tilde F}, ~~d{\tilde H} = \frac{1}{2}\tilde{F}\wedge \tilde{F}
\eea
Similarly, the independent field equations of the near horizon fields are as follows. The 2-form field equation (\ref{2feq}) gives,
\bea
\label{feq1}
\tilde{\nabla}^\ell{(e^{\frac{\Phi}{2}} {\tilde F}_{i \ell})} - e^{\frac{\Phi}{2}}{\tilde F}_{i \ell}h^{\ell} - e^{\frac{\Phi}{2}}T_{i} - e^{\Phi}L_{i}\alpha + \frac{1}{2}e^{\Phi}{\tilde F}^{\ell_1 \ell_2}{\tilde H}_{i \ell_1 \ell_2} = 0
\eea
the 3-form field equation (\ref{3feq}) gives,
\bea
\label{feq2}
\tilde{\nabla}^\ell{(e^{\Phi} L_\ell)}= 0
\eea
and
\bea
\label{feq3}
\tilde{\nabla}^\ell{(e^{\Phi} {\tilde H}_{i j \ell})} - e^{\Phi}h^{\ell}{\tilde H}_{i j \ell} + e^{\Phi}M_{i j} = 0
\eea
The $+-$ and $ij$-component of the Einstein equation (\ref{eins}) gives
\bea
\label{feq4}
- \Delta - \tfrac{1}{2}h^2 + \tfrac{1}{2}\tilde{\nabla}^i(h_i) &=& \tfrac{1}{2}e^{\frac{\Phi}{2}}\bigg(-\tfrac{3}{4}\alpha^2 - \tfrac{1}{8}{\tilde F}^2 \bigg)
\nonumber\\
&+& \tfrac{1}{4}e^{\Phi}\bigg(-L^2 - \tfrac{1}{6}{\tilde H}^2\bigg) + 2g^2 e^{-\frac{\Phi}{2}}
\eea
and
\bea
\label{feq5}
\tilde{R}_{i j} &=& -\tilde{\nabla}_{(i}h_{j)} + \frac{1}{2}h_i h_j
+ \frac{1}{2}e^{\frac{\Phi}{2}}\bigg({\tilde F}_{i \ell}{\tilde F}_{j}{}^{\ell} - \frac{1}{8}{\tilde F}^2 \delta_{i j}\bigg)
\nonumber\\
&&+ \frac{1}{8}e^{\frac{\Phi}{2}}\alpha^2 \delta_{i j}
\nonumber \\
&+& \frac{1}{4}e^{\Phi}\bigg({\tilde H}_{i \ell_1 \ell_2}{\tilde H}_{j}{}^{\ell_1 \ell_2} - \frac{1}{6}{\tilde H}^2 \delta_{i j}\bigg)
\nonumber\\
&&+ \frac{1}{4}e^{\Phi}\bigg(-2L_{i}L_{j} + L^2 \delta_{i j}\bigg)
+ 2g^2 e^{-\frac{\Phi}{2}} \delta_{i j}
\eea
The scalar field equation (\ref{scalarfeq}) gives
\bea
\label{feq6}
{\tilde \nabla}^{i}{{\tilde \nabla}_{i}}{\Phi} - h_{i}{\tilde \nabla}^i{\Phi} = - \frac{1}{2}e^{\frac{\Phi}{2}}\alpha^2 + \frac{1}{4}e^{\frac{\Phi}{2}}{\tilde F}^2 - e^{\Phi}L^2 + \frac{1}{6}e^{\Phi}{\tilde H}^2 - 8g^2 e^{-\frac{\Phi}{2}}
\eea
We remark that the $++$ and $+i$ components of the
Einstein equations are given by
\bea
\label{feq7}
&&\tfrac{1}{2} \tn^2 \Delta -\tfrac{3}{2} h^i \tn_i \Delta -\tfrac{1}{2} \Delta \tn^i h_i + \Delta h^2 +\tfrac{1}{4} (dh)_{ij} (dh)^{ij}
\nonumber\\
&&\quad= \tfrac{1}{2}e^{\frac{\Phi}{2}}T^{i}T_{i} + \tfrac{1}{4}e^{\Phi}M^{i j}M_{i j}
\eea
and
\bea
\label{feq8}
\tfrac{1}{2} \tn^j (dh)_{ij} - (dh)_{ij} h^j - \tn_i \Delta + \Delta h_i
&=& \tfrac{1}{2}e^{\frac{\Phi}{2}}(-\alpha T_i + T^{j}{\tilde F}_{i j})
\nonumber \\
&+& \tfrac{1}{4}e^{\Phi}(-2L_{j}M_{i}{}^{j} + M^{j k}{\tilde H}_{i j k})
\eea
These are implied by (\ref{feq1}), (\ref{feq2}), (\ref{feq3}), (\ref{feq4}), (\ref{feq5}) and (\ref{feq6})
and the Bianchi identities (\ref{beq}) and (\ref{beq2}).

\bibliography{d=6_journal}{}
\bibliographystyle{unsrt}

\end{document}